\documentclass[a4paper,11pt]{article}

\usepackage[utf8]{inputenc}
\usepackage[T1,T2A]{fontenc}
\usepackage{amsmath,amsfonts,amssymb}

 \usepackage{graphicx}
\usepackage{wrapfig}
\usepackage{subfig}
\usepackage{hyperref}
\usepackage[dvipsnames]{xcolor}

\begin{document}

\begin{titlepage}

\begin{center}
{\LARGE {\bf  Cosmological gravitational particle production: Starobinsky vs Bogolyubov, 
 uncertainties, \\ and issues }} \\
\vspace{2cm}
{\bf  Duarte Feiteira and 
Oleg Lebedev$^{}$}
\end{center}

\begin{center}
  \it{Department of Physics and Helsinki Institute of Physics,\\
  Gustaf H\"allstr\"omin katu 2a, FI-00014 Helsinki, Finland}\\
\end{center}

\vspace{2.5cm}

\begin{center} {\bf Abstract} \end{center}
\noindent
 We study  production of free and feebly interacting scalars during inflation using the Bogolyubov coefficient and Starobinsky stochastic approaches.
While the two methods agree in the limit of infinitely long inflation, the Starobinsky approach is more suitable for studying realistic situations, where
the duration of inflation is finite and the scalar field has non-trivial initial conditions.
We find that the abundance of produced particles is sensitive to pre-inflationary   initial conditions, resulting in the uncertainty of many orders of magnitude.
Nevertheless, a lower bound on the particle abundance can be obtained. 
High scale inflation is very efficient in particle production, which leads to strong constraints on the existence of stable scalars 
  with masses below the inflationary Hubble rate. For example, free stable scalars are allowed only if they have masses below an eV or  the reheating temperature is in the GeV range or below.
  We find  universal scaling behavior of the particle abundance, which covers free and feebly interacting scalars as well as those 
  with a small non-minimal coupling to gravity. These considerations are important in the context of non-thermal dark matter since 
   inflationary particle production provides an irreducible background  for other production mechanisms.

\end{titlepage}

\vspace{2.5cm}

\tableofcontents

\section{Introduction}

Particle production in the Early Universe is one of the cornerstones of modern cosmology.
First and foremost, it is responsible for reheating the Universe after inflation \cite{Mukhanov:2005sc}.
It is also important in the context of dark matter as well as   stable or long-lived dark relics. In particular, if
 their particle number is approximately conserved,
they survive up to the present day. This is the case for free or very weakly interacting particles. Despite the absence 
of any significant couplings, such particles  can  copiously be produced by classical gravity
during inflation. The observational bounds on the dark relic abundance provide us with an important probe of the Early Universe dynamics as well as of 
the underlying fundamental theory.

Various aspects of gravitational particle production have  been studied since 1960's \cite{Parker:1969au,Parker:1971pt,Grib:1976pw,Ford:1986sy}. Recent reviews of the subject can be found in 
  \cite{Ford:2021syk,Kolb:2023ydq}.
Typically,  gravitational particle production is understood as particle production in  time-dependent spaces whose dynamics is driven by gravity. The prime example of such a time-dependent background
is provided by 
 cosmological inflation      \cite{Starobinsky:1980te,Guth:1980zm,Linde:1981mu}, which is the focus of 
the present work.  After inflation, the inflaton field oscillates around the minimum of its potential inducing oscillations in the metric and also leading to particle production \cite{Ema:2015dka}.
Gravity effects may  be significant during reheating and 
certain production processes can be attributed to  the $s$-channel graviton exchange \cite{Garny:2015sjg,Mambrini:2021zpp}. 
While 
the classical gravity approximation is expected to break down at high energies, 
 quantum gravitational effects  on particle production 
after inflation 
can be parametrized by a set of higher-dimensional Planck-suppressed operators \cite{Lebedev:2022ljz}.

In our  work, we study the problem of inflationary scalar production from two viewpoints: the Bogolyubov coefficient   \cite{Bogolyubov:1958se}   method, based on the wavefunctions
with  $in$ and $out$ asymptotics, 
  and the Starobinsky  stochastic  approach \cite{Starobinsky:1986fx}, which focuses on the evolution of the long wavelength scalar ``condensate''.
We find that these seemingly very different  approaches lead to the same particle abundance in the limit of infinitely long inflation. The Starobinsky approach, on the other hand, 
is more adaptable to realistic situations with finite duration of inflation. We study the dependence of the particle abundance  on pre-inflationary  initial conditions and derive
constraints on stable spin-0 particles in inflationary frameworks. Our findings indicate that the existence of free or feebly coupled stable scalars is  compatible with standard high scale inflation only  
under special circumstances, e.g. if the reheating temperature is very low or the scalars are very light. These results extend and improve the earlier analysis \cite{Lebedev:2022cic}.

\section{The Bogolyubov coefficient approach}

The Bogolyubov coefficient  method is based on the time-dependent particle number operator $N$. 
The initial state     $| 0^{\rm in} \rangle$   has no particles according to the  $in$-definition $N^{\rm in}$, whereas
it contains particles with respect to the late time number operator $N^{\rm out}$.
The two definitions are related by a Bogolyubov transformation, hence the name.

\subsection{Basics}

In what follows, we adapt  the conventions and notation of Ref.\,\cite{Kolb:2023ydq}.
Consider a real scalar $\Phi$ with the 
  Lagrangian  
\begin{equation}
{\cal L}={1\over 2}g^{\mu \nu} \partial_\mu \Phi \, \partial_\nu \Phi - V(\Phi) +{1\over 2} \xi \Phi^2 R\;,
\end{equation}
where $R$ is the scalar curvature,
$\xi$ is the non-minimal coupling to gravity \cite{Chernikov:1968zm}, and 
  the potential for a free scalar  is given by 
\begin{equation}
V(\Phi) = {1\over 2} m^2  \Phi^2 \;.
\end{equation}
The potential will later be extended to include a small self-interaction term. The Friedmann metric in terms of the conformal time variable $\eta$ is
$g_{\mu\nu}= a^2 (\eta) \, {\rm diag} (1,-1,-1,-1)$. The cosmological time $t$ is related to the conformal time by $dt = a\, d\eta$, while  the Hubble
rate and the curvature are  $H(\eta)= a^\prime (\eta)/a^2(\eta)$ and $R(\eta)= -6 a^{\prime\prime}/a^3$, respectively. 
We focus on light fields such that $m\ll H$ during inflation.

It is convenient to define a rescaled field variable
\begin{equation}
\phi = a(\eta) \, \Phi \;,
\end{equation}
which can  further be decomposed in spacial Fourier modes,
\begin{equation}
\phi (\eta, {\bf x})= \int {d^3 {\bf k} \over (2\pi)^3}
\left[ a_{\bf k}\, \chi_k (\eta)\, e^{i {\bf k} \cdot {\bf x} } +  a_{\bf k}^\dagger \, \chi_k^* (\eta)\, e^{-i {\bf k} \cdot {\bf x} }
\right]\,.
\label{fourier}
\end{equation}
Here $a_{\bf k}, a_{\bf k}^\dagger $ are the annihilation and creation operators with the usual commutation relations,
$\left[   a_{\bf k}  ,    a_{\bf q}^\dagger   \right]= (2\pi)^3 \delta ({\bf k} - {\bf q})$ and the other commutators being zero; $k \equiv |{\bf k} |$.
The equation of motion for $\Phi$ translates into the  $\chi_k (\eta)$ mode  equations
 \begin{equation}
    \chi_k^{\prime\prime} + \omega_k^2 \, \chi_k=0 \;,
    \label{mode-eq}
    \end{equation}
with 
\begin{equation}
    \omega_k^2 (\eta) = k^2 + a^2(\eta)\, m^2 + \left(   {1\over 6} - \xi  \right)\, a^2(\eta)\, R(\eta) \;.
    \label{general-omega}
    \end{equation}

The scale factor satisfies $a(\eta \rightarrow -\infty) \rightarrow 0$ and $a(\eta \rightarrow +\infty) \rightarrow \infty$ with $H (\eta \rightarrow +\infty) \rightarrow 0$.
In these asymptotic regimes,   the space is effectively flat and one may use the flat space mode functions.
Since the frequency changes adiabatically at early and late times,  one can define the $in$ and $out$ solutions with the boundary conditions,
\begin{eqnarray}
&& \chi_k^{\rm in} (\eta \rightarrow -\infty) \rightarrow {e^{-i \int^\eta d \eta' \omega_k (\eta') } \over \sqrt{2 \omega_k (\eta)}} \;, \\
&& \chi_k^{\rm out} (\eta \rightarrow +\infty) \rightarrow {e^{-i \int^\eta d \eta' \omega_k (\eta') } \over \sqrt{2 \omega_k (\eta)}} \; .
\end{eqnarray}
Since the differential equation is of second order, these solutions are not independent. They are related by a linear transformation with $constant$ coefficients,
\begin{equation}
  \chi_k^{\rm in} (\eta ) = \alpha_k \, \chi_k^{\rm out} (\eta ) + \beta_k \, \chi_k^{\rm out\, *} (\eta ) \;.
 \end{equation}
This change of the basis of the mode functions corresponds to the change in the definition of the creation and annihilation operators such that the combination 
$a_{\bf k}\, \chi_k (\eta)  +  a_{-\bf k}^\dagger \, \chi_k^* (\eta)$, which defines $\phi (\eta, {\bf x})$,      remains invariant.\footnote{This expression appears in (\ref{fourier}) if the integration variable is changed to $-{\bf k}$ in the second term.}
 The transformation belongs to the SL(2,C) group, i.e. 
 \begin{equation} 
 |\alpha_k|^2 -  |\beta_k|^2=1\;,
\end{equation}
and 
\begin{equation} 
  a_{\bf k}^{\rm in} = \alpha_k^*  \, a_{\bf k}^{\rm out} - \beta_k^* \, a_{-\bf k}^{\rm out\, \dagger}\;.
\end{equation}
This implies, in particular, that the basis change (Bogolyubov)  coefficients  $\beta_k$    can be found via
\begin{equation} 
 \beta_k = i (\chi_k^{\rm out\,\prime}    \chi_k^{\rm in}  - \chi_k^{\rm out}    \chi_k^{\rm in\,\prime} )\;.
 \label{beta-def}
 \end{equation}

The operators $a_{\bf k}^{\rm in} $ are used to define the vacuum state $| 0^{\rm in} \rangle $  in the infinite past and construct the Fock space via
 \begin{equation} 
  a_{\bf k}^{\rm in} \, | 0^{\rm in} \rangle =0\;.
 \end{equation}
This is known as the 
{\it Bunch-Davies} vacuum \cite{Bunch:1978yq} in curved space, corresponding to the flat space vacuum at early times.
In the Heisenberg picture, this state is time-independent, while the operators, including the particle number operator,
are time-dependent. 
The $in$-vacuum may contain particles according to the definition of a particle based on the $out$-vacuum. The $out$ number operator is 
\begin{equation} 
N^{\rm out} =  \int {d^3 {\bf k} \over (2\pi)^3}\,  a_{\bf k}^{\rm out\, \dagger} a_{\bf k}^{\rm out}\;.
 \end{equation}
The particle number in the $in$-vacuum is then
\begin{equation} 
\langle 0^{\rm in}  | N^{\rm out} | 0^{\rm in} \rangle = V \, \int {d^3 {\bf k} \over (2\pi)^3}\,   |\beta_k|^2\;,
 \end{equation}
where $V$ is the space volume factor appearing due to $\delta ({\bf 0})$. The physical particle density is given by $n = {1\over V} \langle 0^{\rm in}  | N^{\rm out} | 0^{\rm in} \rangle $, which means that the 
$comoving$ particle density $a^3 n$    is 
\begin{equation} 
a^3 n =  \, \int {d^3 {\bf k} \over (2\pi)^3}\,   |\beta_k|^2 \equiv \int {dk \over k} \, a^3 n_k ~~{\rm with }~~ a^3 n_k \equiv {k^3 \over 2\pi^2 } \, |\beta_k|^2\;.
 \end{equation}
This is the basis for the particle production computations.

It is important to remember that this result is based on the Bunch-Davies vacuum, i.e. absence of any $in$-particles,  in the infinite past ($\eta \rightarrow -\infty$). 
Although this assumption is  appropriate for most purposes, its scope of applicability is nevertheless limited: in reality, the initial state may be prepared at finite
$\eta<0$ and with a non-zero particle number. The Bogolyubov coefficient method is general and can be adapted to such cases, yet this is not the conventional 
approach.

\subsection{Calculational  procedure}

To obtain  the particle density, one needs to compute the Bogolyubov coefficient $\beta_k$  via (\ref{beta-def}). Given the expanding background $a(\eta)$, one computes the 
frequency function $\omega_k$ in (\ref{general-omega}) and  solutions to (\ref{mode-eq}) with the $in$ and $out$ boundary conditions.

The Bogolyubov coefficient is constant,
\begin{equation}
\beta^\prime_k (\eta) =0 \;,
\end{equation}
which allows for its computation at any convenient point $\eta$.
The mode equation can be solved analytically separately in two regimes: inflation ($\eta <0$), with the $in$ boundary condition,  and radiation (or matter) domination  epoch ($\eta >0$), with the $out$ 
boundary condition.
The Bogolyubov coefficient is then computed at $\eta \sim 0$, where both solutions are  valid approximately. 

We choose  a smooth  function $a(\eta)$ which interpolates between the inflationary and radiation/matter domination regimes. 
Its second derivative is, however, discontinuous at the origin $\eta=0$, leading to a jump in the curvature $R$:
\begin{equation}
R= - {6a^{\prime\prime} \over a^3} \;.
\end{equation}
Nevertheless, the solution $\chi_k (\eta)$ is smooth at the origin and the Hubble rate   $H(\eta)= a^\prime /a^2$     is continuous.

Specifically, the function $a(\eta)$ corresponding to inflation followed by a {\it radiation domination} epoch is taken to be
  \begin{eqnarray}
 && a(\eta) = \left\{  
 \left(  {1\over a_e H_e}  -\eta  \right)^{-1} H_e^{-1}
  ~~{\rm  for }~~ \eta\leq 0 ~~,~~ a^2_e H_e \left( \eta + {1\over a_e H_e}\right) ~ ~{\rm for}~ \eta>0 \right\}~,\\
 && H(\eta) = \left\{H_e ~~{\rm for}~~ \eta \leq 0 ~~,~~ H_e \, (a_e/a)^2 ~ ~{\rm for}~ ~\eta>0 \right\}~,
 \end{eqnarray}
where $a_e$ and $H_e$ are the scale factor and the Hubble rate at the end of inflation, respectively.
For the {\it matter domination} case, the $\eta >0 $ branch is replaced by
    \begin{equation}
 H=H_e \left( {a_e\over a}\right)^{3/2} ~~,~~ a= {1\over 4} a_e^3 H_e^2 \left( \eta + {2\over a_e H_e}   \right)^2 \;.
 \end{equation}
The curvature jumps from $R=-12H_e^2$ at $\eta\leq 0$ to $R=0$ at $\eta >0 $ in the radiation domination case,
while, for the matter domination period, $R\simeq -3H_e^2$ at small $\eta >0$.

Our inflationary solution $\chi_k^{\rm in} (\eta)$ uses the approximation $a\propto 1/\eta$ and hence 
is valid for $\eta \lesssim -{1 \over a_e H_e}$. The $out$ solutions, on the other hand, apply at $\eta \gtrsim { 1 \over a_e H_e}$. 
 Hence, in order to compute the Bogolyubov coefficient at a given point, e.g. $\eta_0 \simeq { 1 \over a_e H_e}$, it is necessary 
 to extrapolate the $in$ solution to small positive values of $\eta$.

Inspection of the above exact $a(\eta)$ functions shows that at small $|\eta| < { 1 \over a_e H_e}$, the scale factor is approximately constant and $R$
is constant piecewise.  The EOM solutions in this region are therefore of the form $e^{\pm \omega \eta}$ or $e^{\pm i \omega \eta}$, with 
approximately constant frequency and  appropriate boundary conditions at $ \eta \sim -{1 \over a_e H_e}$. In the regime of interest, $\omega \sim a_e H_e$ or smaller, hence the wave function variation over the period 
$\Delta \eta \sim {1 \over a_e H_e}$   is ${\cal O}(1)$. This evolution effectively ``rotates'' the wave function and its first derivative. Thus, for small enough momenta, 
the  size of the Bogolyubov coefficient can be estimated analytically as
\begin{equation}
|\beta_k| \simeq  \kappa \; {\rm max }\{  \, |\chi_k^{\rm in }(-\eta_0)  \chi_k^{\rm out\prime } (\eta_0)| \; , \; |   \chi_k^{\rm in \prime} (-\eta_0) \chi_k^{\rm out} (\eta_0)  |  \,  \} \:,
\label{beta-approx}
\end{equation}
with $\eta_0 \sim { 1 \over a_e H_e}$ and $\kappa $ parametrizing the extrapolation uncertainty, $\kappa \sim {\cal O}(1)$. 

Note that this approximation breaks down at  $k \gtrsim a_eH_e$. Indeed, at large momenta, $\omega_k \sim k $ and $\beta_k $ vanishes due to a cancellation between the two terms.
This corresponds to the flat space limit and no particle production. For superheavy dark relics       \cite{Chung:1998zb,Chung:1998ua,Kuzmin:1998kk}, $m \gtrsim H$, the above approximation is also inadequate.  

In our numerical analysis, we do not resort to    approximations and calculate $\beta_k$ as defined by (\ref{beta-def}).

\subsection{Inflation}

The mode equation  for a  free scalar during inflation is 
 \begin{equation}
    \chi_k^{\prime\prime} + \omega_k^2 \chi_k=0
    \end{equation}
   with 
    \begin{equation}
   \omega_k^2 = k^2 + {1\over \eta^2} \, ({m^2/H_e^2-2+12\xi}) \;
    \end{equation}
at $\eta \ll -{1\over a_e H_e }$
and we assume a constant Hubble rate $H=H_e$ during inflation.
The solution is subject to the $in$ initial condition,
 \begin{equation}
    \chi_k^{\rm in} (\eta \rightarrow - \infty) \rightarrow {1\over \sqrt{2k}}\, e^{-ik\eta}\;.
    \end{equation}

The above equation is of Bessel type and the solution is
 \begin{equation}
    \chi_k^{\rm in} (\eta ) = \sqrt{-\pi \eta \over 4} H_\nu^{(1)} (-k\eta) \times {\rm phase} \;,
  \label{hankel}
    \end{equation}
with a constant ``phase'', which is irrelevant for our calculation, and 
 \begin{equation}
  \nu = \sqrt{{9\over 4} - {m^2 \over H_e^2}-12\xi} \simeq {3\over 2} - \delta \;,
    \end{equation}
where  $\delta \equiv  {1\over 3} \, {m^2 \over H_e^2} \ll 1$ and $\xi=0$.

At the end of inflation, $ |k\eta | \ll 1$ and the wave function can be approximated by 
 \begin{equation}
   \chi_k^{\rm in} (\eta ) \simeq {1\over \sqrt{2} \,k^{3/2-\delta}} \, {1\over \eta^{1-\delta}} \times {\rm phase}\;.
   \label{in-appr}
    \end{equation}

 \subsection{Radiation domination}

During the radiation era, $R=0$ and 
     \begin{equation}
   \omega_k^2 = k^2 + a^2(\eta) m^2 \;.
    \end{equation}
    At $\eta > {1\over a_e H_e }$, one can approximate
   \begin{equation}
  a(\eta) \simeq a_e^2 H_e \eta~~,~~ H(\eta) \simeq {1\over a_e^2 H_e }\, {1\over \eta^2}\;.
    \end{equation}  
   The above equation is of the form
  $$ y^{''} (z) + (z^2 +\lambda) \, y (z)=0 \;,$$
   which is solved by the parabolic cylinder function
     $$  y(z)= D_{- {1+i\lambda \over 2}} \left[  \pm (1+i) z  \right] \, .  $$
     In our case,
     \begin{equation}
     y=   ma_e^2 H_e \;    \chi_k ~~,~~ z= \sqrt{ ma_e^2 H_e}\;\eta ~~,~~ \lambda={k^2 \over ma_e^2 H_e}\;.
     \end{equation}
     The solution is subject to the $out$ boundary condition,
      \begin{equation}
    \chi_k^{\rm out} (\eta \rightarrow \infty) \rightarrow {e^{-i \,{1\over 2} m a_e^2 H_e \eta^2} \over \sqrt{2ma_e^2 H_e \eta}} \;,
     \end{equation}
     since $\omega_k \rightarrow ma_e^2H_e\eta$.
     The asymptotic behavior at large arguments of the parabolic cylinder function is given by 
     $$         D_\nu (z)  \rightarrow e^{-z^2/4} \, z^\nu  \;.   $$
     Therefore, the correct solution is 
      \begin{equation}
    \chi_k^{\rm out} (\eta ) =  {e^{-\pi C/4}    \over  (2ma_e^2 H_e)^{1/4}    }\, D_{-iC-1/2} \left(    e^{i\pi/4} \sqrt{2ma_e^2 H_e} \, \eta  \right)    \times {\rm phase}   \;,
     \end{equation}
   where
      \begin{equation}
  C = {k^2 \over 2ma_e^2 H_e} 
       \end{equation}
    and the ``phase'' is an irrelevant constant phase.\footnote{The asymptotic of the $D$-function also involves the $\eta$-dependent  phase proportional to $\ln \eta /\eta^2 $, which vanishes at
    $\eta \rightarrow \infty$.   } 
     
     To compute the Bogolyubov coefficient, we need the early time behavior of  $\chi_k^{\rm out} (\eta ) $ at the beginning of the radiation epoch.  Introduce
     \begin{equation}
     k_* = \sqrt{ma_e^2 H_e} \;,
     \end{equation}
     which satisfies
     $$   {k_* \over a_e H_e} \ll 1\;.   $$
     The beginning of the radiation epoch corresponds to $\eta \sim {1\over a_e H_e}$, which implies that the argument of the $D$-function is small in this regime and one can use the corresponding Taylor series,
     \begin{equation}
     D_{-iC-1/2} (z) \simeq {    \sqrt{\pi }    \, 2^{-iC/2 -1/4} \over \Gamma (iC/2 +3/4) } - z\; {    \sqrt{\pi }    \, 2^{-iC/2 +1/4} \over \Gamma (iC/2 +1/4) } \;.
     \end{equation}
     Consider first the small momentum regime
     $$    k \ll k_* \;. $$
     In this case, $C \ll 1$ and 
      \begin{equation}
    \chi_k^{\rm out} (\eta ) \simeq {1\over (2 k_*^2)^{1/4}} \, \left(     { \sqrt{\pi }    \, 2^{-1/4}   \over \Gamma (3/4) }      -       { \sqrt{\pi }    \, 2^{1/4}   \over \Gamma (1/4) }   \, e^{i\pi/4} \sqrt{2} k_* \eta   \right)   
         \end{equation}
   up to a constant phase   in this regime. Note that the result is $k$-independent.
     
     Evaluating the Bogolyubov coefficient at $\eta \sim {1\over a_e H_e}$, one finds
      \begin{equation}
  | \chi_k^{\rm in \prime}  \chi_k^{\rm out} | \gg   | \chi_k^{\rm in }  \chi_k^{\rm out\prime } |
           \end{equation}
     due to $k_* \eta \ll 1$.
     Since 
     $$ | \chi_k^{\rm in \prime} | \simeq {1\over \sqrt{2} k^{3/2-\delta}} \, {1\over \eta^{2-\delta}} $$
      with $\delta \equiv {1\over 3} {m^2\over H_e^2}$,
      \begin{equation}
 |\beta_k|^2\simeq \kappa^2 \; {(a_e H_e)^{4-2\delta} \over k^{3-2\delta} k_*} \;, 
           \end{equation}
     where the constant $\kappa $ is ${\cal O}(1)$ and parametrizes the uncertainty in the wave function extrapolation, see Eq.\,\ref{beta-approx}.

         For $k> k_*$, the wave function $ \chi_k^{\rm out} (\eta ) $ decreases with $k$ at early times, hence the spectrum gets suppressed and does not significantly contribute to the total particle number, while, at very large $k$, $\beta_k \simeq 0$ 
     since the wave functions become plane waves.
   The physics of the cutoff $k_*$ can be understood as follows. Since $H= H_e (a_e/a)^2$, the Hubble rate approaches the particle mass (from above) at 
   \begin{equation}
   a_m = a_e \, \sqrt{H_e\over m}
   \label{am}
   \end{equation}
   and 
    \begin{equation}
   {k_*\over a_m} =  m \;.
   \end{equation}
   The left hand side then represents  the maximal physical momentum that can be excited at the time when $H=m$. Therefore, all the created particles are non- or semi-relativistic at $a=a_m$. 
   
\begin{figure}[h!]
    \centering
    \includegraphics[width=0.49\textwidth]{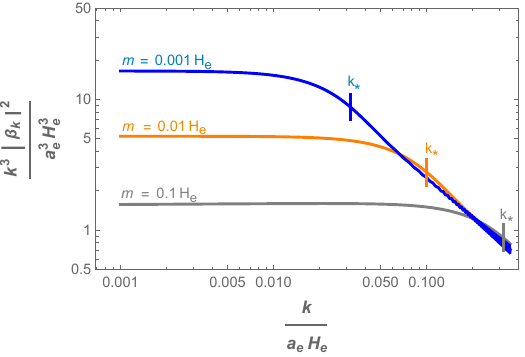}
    \includegraphics[width=0.50\textwidth]{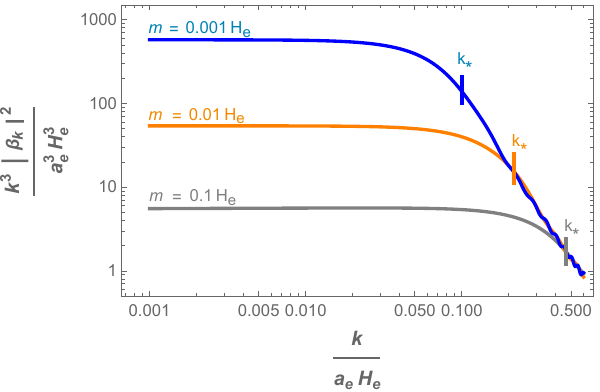}
    \caption{ Normalized Bogolyubov coefficient squared as a function of the momentum, based on a numerical solution with a smooth $a(\eta)$. $k_*$ marks the comoving momentum (in units of $a_eH_e$) beyong which particle production is suppressed.
    {\it Left}: inflation followed by a radiation-dominated era. {\it Right}: inflation followed by a matter-dominated era.}
    \label{beta-plot}
\end{figure}

   Our numerical results with the smooth $a(\eta)$ function, without resorting to simple approximations,  are shown in Fig.\,\ref{beta-plot}. They agree well with our analytical estimates,  in particular, 
   the $\delta$ dependence of
       the particle number with a given momentum $k < k_*$,
   \begin{equation}
  a^3 n_k = \kappa^2 \, {(a_e H_e)^{4} \over 2\pi^2  k_*^{} } \, \left({k\over a_e H_e}\right)^{2\delta}\;.
      \end{equation}

 The total particle number is proportional to the integral 
     $$    \int^{k_*}_0 k^{2\delta -1}  \;,$$
    which is finite and dominated by the momenta of order $k_*$.
    The result is 
    \begin{equation}
 a^3 n = {3\kappa^2 \, a_e^3\over 4\pi^2} \, {H_e^{11/2 }\over  m^{5/2}} \;,
 \label{a3-n-rad}
    \end{equation}
    to leading order in $m/H_e$, where $\kappa \sim {\cal O}(1)$. 
    Note that the particle number diverges in the limit $\delta \rightarrow 0$.\footnote{Refs.\,\cite{Kolb:2023ydq,Jenks:2024fiu} take the limit $\delta \rightarrow 0$, which results in a logarithmic sensitivity to the IR cutoff.}

   \subsection{Non-minimal coupling $|\xi | \ll 1$} 
   \label{sub-xi}
   
   A small value of $\xi$ can be induced by quantum corrections \cite{Buchbinder:2017lnd} and thus is of interest.
   In fact, one normally expects 
   \begin{equation}
   {|\xi | } \gg {m^2 \over H_e^2}
   \end{equation}
   unless the scalar is super-heavy. Then, the inflationary $in$ wave function remains of the form (\ref{hankel}) with
   \begin{equation}
   \nu \simeq {3\over 2} -\delta = {3\over 2} - { 4} \, \xi \;.
   \end{equation}
   The approximation (\ref{in-appr}) also applies with $\delta$ defined above. The $out$ wave function is not modified at all due to $R=0$ during the radiation dominated epoch.
   Therefore, the previous calculations  are straightforwardly adapted to the $\xi \not= 0$ case. The difference appears at the last step of momentum integration and the sign of $\xi$ 
   leads to two distinct results.
   
   \subsubsection{$\xi >0$ }
   
   This creates an effective  positive mass term during inflation and  
    \begin{equation}
 a^3 n = {\kappa^2 \, a_e^3\over 16\pi^2} \, {H_e^{7/2 }\over  m^{1/2}} \, {1\over \xi} \;,
    \end{equation}
   to leading order in $\xi$.
    The resulting particle number is much smaller than that in the $\xi=0$ case, by the factor ${H_e^2 \over m^2} \, \xi$. We note that $\xi$ creates an effective mass during inflation, $m^2_{\rm eff} = 12\xi \,H^2 $, hence
    one naturally obtains (\ref{a3-n-rad}) by replacing $\xi \rightarrow {1\over 12} {m^2 \over H^2}\,$. On the other  hand, the non-minimal coupling to gravity does not behave as a mass term after inflation, so 
    one cannot replace the full $m^2$-dependence with  that of $\xi$.
   
    \subsubsection{$\xi <0$ }
   
   A negative $\xi$ creates an additional tachyonic mass term which amplifies production of the long wavelength modes. The integral over $k$ formally diverges and thus requires an IR cutoff. The result is
    \begin{equation}
 a^3 n = {\kappa^2 \, a_e^3\over 16\pi^2} \,  {H_e^{7/2 }\over  m^{1/2}}  \,   \left(  {H_e\over k_{\rm min}} \right)^{8|\xi|}  \,  {1\over |\xi|} \;,
   \label{neg-xi}
    \end{equation}
   Although the particle number depends on the cutoff, this dependence is weak at small $|\xi| < 10^{-3}$.

    \subsection{Matter domination}
   
   At $\eta >0$, the scale factor and the Hubble rate are given by
    \begin{equation}
 H=H_e \left( {a_e\over a}\right)^{3/2} ~~,~~ a= {1\over 4} a_e^3 H_e^2 \left( \eta + {2\over a_e H_e}   \right)^2 \;.
 \end{equation}
The curvature at $\eta \gg {1\over a_e H_e}$ is 
\begin{equation}
R= -3H^2
\end{equation}
   such that the $\chi_k$ oscillation frequency squared is
    \begin{equation}
    \omega_k^2= k^2 + {1\over 16} a_e^6 H_e^4 m^2 \, \eta^4 - {2\over \eta^2} \;,
    \end{equation}
    setting $\xi=0$.
    
    Consider first long wavelength modes,
    \begin{equation}
    k < k_* = a_e m^{1/3} H_e^{2/3} \;.
    \end{equation}
    In this case, $k^2$ can be dropped from $\omega^2_k$. Indeed, at small $\eta$, the frequency is dominated by the last term, while at large $\eta$, the second term dominates. These two terms become comparable at
    $\eta \sim 1/k_*$ such that, for $k<k_*$, the momentum dependence effectively disappears.

   The consequent equation of motion is of the type
   $$ x^2 y^{\prime\prime} +(b x^n +c)y=0 \;,$$
   which is solved by  \cite{diff-eq}
   $$     y(x)= \sqrt{x} \left[ C_1\, H^{(1)}_\nu \left(   {2\over n}  \sqrt{b} x^{n/2} \right) + C_2\, H^{(2)}_\nu \left(   {2\over n}  \sqrt{b} x^{n/2} \right)
   \right]     \;,        $$
   with $\nu = {1\over n}   \,\sqrt{1-4c}$.
   In our case,
   $$ c=-2 ~,~ b = {1\over 16} \,k_*^6 \;.$$
   
   The correct $out$ asymptotic behaviour is
   \begin{equation}
   \chi_k^{\rm out} (\eta \rightarrow \infty) \rightarrow {e^{  {-i } \,{1\over 12} k_*^3 \eta^3     }    \over \sqrt{ k_*^3 \eta^2/2}} \;,
   \label{out-2}
   \end{equation}
   such that the solution is given by
   \begin{equation}
   \chi_k^{\rm out}  (\eta) = \sqrt{\pi\eta \over 12} \, H^{(2)}_{1/2} \left(     {1\over 12} k_*^3 \eta^3  \right) \times {\rm phase} = {\sqrt{2}    \over k_*^{3/2} \eta} \, e^{  {-i } \,{1\over 12} k_*^3 \eta^3     }  \times {\rm phase}  \;,
   \end{equation}
where ``phase'' stands for a constant irrelevant phase factor.
   Note that, in this case,  $\chi_k^{\rm out} $ happens to coincide with its asymptotic (\ref{out-2}).
   
   At small values of the argument, $k_* \eta \ll 1$,
   \begin{equation}
   \chi_k^{\rm out}  (\eta) \simeq  {\sqrt{2}    \over k_*^{3/2} \eta} \,,
      \end{equation}
   up to a constant phase. This approximation is legitimate at $\eta \sim 1/(a_e H_e)$ as long as $m \ll H_e$.
   Using the $in$ solution described earlier, we get at $|\eta| \sim 1/(a_e H_e)$,
  \begin{equation}
  | \beta_k | = \kappa \, {(a_e H_e)^{3-\delta} \over k^{3/2-\delta} k_*^{3/2} } \;,
      \end{equation} 
   where $\kappa$ is an order one constant associated with the wave function extrapolation from $\eta \sim -1/(a_e H_e) $ to $\eta \sim 1/(a_e H_e) $.
   We note that, if  $\chi_k^{\rm in} \propto 1/\eta^{1-\delta}$ is taken  to be valid literally at $\eta >0$, there is a strong cancellation in $\beta_k$ according to  (\ref{beta-def}).
   However, this effect is spurious since the wave function evolution through $\eta=0$ affects it and its derivative differently, spoiling the cancellation. 
   
   The particle number with a given momentum $k < k_*$ is 
   \begin{equation}
  a^3 n_k = \kappa^2 \, {(a_e H_e)^{6} \over 2\pi^2  k_*^{3} } \, \left({k\over a_e H_e}\right)^{2\delta}\;.
  \label{a3-nk-2}
      \end{equation}
   For larger $k> k_*$, the quantity $\omega_k^2$ becomes less tachyonic at small $\eta$, therefore $\chi_k^{\rm out}$ decreases with $k$ (see Fig.\,\ref{beta-plot}, right panel). The resulting contribution to the total particle number is small and, neglecting it, we get
   \begin{equation}
  a^3 n \simeq \int^{k_*}_0 {dk\over k}\, a^3 n_k = {3\kappa^2 \over 4\pi^2 }\,a_e^3\, {H_e^6\over m^3}\;.
     \label{a3-n-mat}
        \end{equation}
Again, a non-zero $\delta$ is necessary to obtain a finite particle number.

   \subsection{Non-minimal coupling $|\xi | \ll 1$}

  A small non-minimal coupling to gravity  is included in a straightforward manner, as before.   The frequency $\omega_k$  remains essentially unaffected, while 
  the $out $ wave function is independent of $k$ for $k<k_*$.
   Hence, 
  in $\chi_k^{\rm out}$, a small $|\xi | \ll 1$ can be neglected.
   In the $in$ wave function, on the other hand,  it affects the momentum dependence, as in the radiation domination case. 
   The calculation parallels that in the radiation domination case and the  results are as follows.
   
   \subsubsection{$\xi >0$ }
  
    \begin{equation}
 a^3 n = {\kappa^2 \, a_e^3\over 16\pi^2} \, {H_e^{4 }\over  m^{}} \, {1\over \xi} \;.
    \end{equation}
As in the radiation domination case, replacing  the non-minimal coupling    $\xi \rightarrow {1\over 12} {m^2 \over H^2}\,$ produces (\ref{a3-n-mat}).

 \subsubsection{$\xi <0$ }
  
     The result is formally divergent and requires a cutoff, as in (\ref{neg-xi}),
    \begin{equation}
 a^3 n = {\kappa^2 \, a_e^3\over 16\pi^2} \,  {H_e^{4 }\over  m^{}}  \,   \left(  {H_e\over k_{\rm min}} \right)^{8|\xi|}  \,  {1\over |\xi|} \;.
    \end{equation}

   \subsection{Field size at the end of inflation}
   \label{Bog-field}
   
   The average field size  is given by the square root of the variance $ \langle \Phi^2       \rangle$.
   Consider the equal time correlator in de Sitter space ($H={\rm const}$)
   \begin{equation}
   \langle 0^{\rm in} \left| \Phi (\eta, {\bf 0}) \Phi (\eta, {\bf r})  \right|  0^{\rm in}        \rangle = {1\over a^2 (\eta) } \, \int {d^3 {\bf k} \over (2\pi)^3} \left| \chi_k^{\rm in } (\eta)  \right|^2 \,  e^{i {\bf k} \cdot {\bf r}}\;,
   \end{equation}
   at  the end of inflation, $\eta \rightarrow -0$, and small spacial separations $|{\bf r}| \rightarrow 0$.
   Following  the Starobinsky approach, let us split the long and short wave modes in the integral. The long wave modes are defined by $|{\bf k}|\equiv k \leq \epsilon a H$, with $\epsilon \ll 1$, while the short wave modes correspond to
  $ k > \epsilon a H$. At late times corresponding to 
  the end of inflation, $|k\eta| \ll 1$, hence the infrared (IR) part of the integral is determined by
  \begin{equation}
 \left| \chi_k^{\rm in } (\eta)  \right| \simeq    {1\over \sqrt{2} \,k^{3/2-\delta}} \, {1\over \left| \eta\right|^{1-\delta} }  \;.
   \end{equation}
   We may set ${\bf r} =0$ in the IR contribution  such that 
   \begin{equation} 
   \langle \Phi^2       \rangle_{\rm IR} =    {1\over a^2 (\eta) } \, \int^{\epsilon aH} {d^3 {\bf k} \over (2\pi)^3} \left| \chi_k^{\rm in } (\eta)  \right|^2 = {3\over 8\pi^2} \,{H^4 \over m^2}   \;,
   \label{var-Phi}
   \end{equation}
where we have used $\eta = -1/(aH)$ and $\epsilon^\delta \rightarrow 1$. The result is finite as $\eta \rightarrow 0$ and represents an anomalously   large correlator, enhanced by $H^2/m^2 \gg 1$ compared to the 
thermal value based on the Gibbons-Hawking temperature $T=H/(2\pi)$. We note that, since what matters in this calculation is $\delta$, the above expression  also applies to the effective mass term induced by a small non-minimal coupling 
to gravity $\xi>0$.

   The UV part of the integral can be  split into an intermediate momentum range $\epsilon a H < k < \Lambda  a H$ and the high momentum range $k > \Lambda a H$ with $\Lambda \gg 1$. 
   To get a sensible answer, one must keep a nonzero spacial separation ${\bf r}$ and integrate over the angular variables, which produces the factor ${\sin k|{\bf r}|  \over k |{\bf r}|  }   $. 
   One can then  show that  the UV piece does not exhibit the $H^2/m^2$ enhancement and 
   behaves    as  
   $1/(a |{\bf r }|)^2  =1/ r^2_{\rm phys} $, where $r_{\rm phys}$ is the physical distance between the two points.\footnote{This is equivalent to the $\eta^2/{\bf r}^2$ behavior such that 
   the UV piece formally vanishes if $\eta \rightarrow 0$ faster than ${\bf r}\rightarrow 0$.}  Although this piece formally diverges as $r_{\rm phys} \rightarrow 0$ and requires renormalization \cite{Bunch:1978yq}, 
   it is negligible compared to the IR contribution  if      the physical distance is small  (well within the Hubble patch) but non-zero   such that  $r_{\rm phys}\gg H^{-1}\, {m\over H}$ with $m/H \ll 1$.
   
   The long wavelength modes are approximately constant over the Hubble patch $H^{-1}$, hence one can define 
   the mean field as 
   \begin{equation}
   \overline{\Phi} \equiv \sqrt{ \langle \Phi^2       \rangle_{\rm IR} }\;.
   \end{equation}
    It can be treated as a ``condensate'' and subsequently be used to compute the particle density produced by inflation.

   In the massless case $\delta=0$, the correlator  diverges logarithmically in the infrared and requires an IR cufoff $k_{\rm min}$. The result is
   \begin{equation} 
   \langle \Phi^2       \rangle_{\rm IR} =       \langle \Phi^2       \rangle_{\rm IR}^0  + {H^2 \over 4\pi^2 } \, \ln a =  \langle \Phi^2       \rangle_{\rm IR}^0        + {H^3 \over 4\pi^2 } \, (t-t_0) \;,
   \end{equation}
   where $t$ is the standard cosmological time and the initial value of the correlator $ \langle \Phi^2       \rangle_{\rm IR}^0 $ is set at $a=1$ or $t=t_0.$
   This is the well known ``random walk'' result: the average field value grows as a square root of time and diverges as $a \rightarrow \infty$. The UV part of the correlator, on the other hand,  retains the properties outlined above.

   \section{The Starobinsky approach}
   
   The Starobinsky stochastic approach \cite{Starobinsky:1980te}    is based on the evolution of the long wavelength field modes, which define an approximately constant  field within a Hubble patch.  After inflation, this field or a ``condensate'' can be converted into   
   a particle density, thus providing us with another view on inflationary particle production. A pedagogical exposition of the stochastic approach can be found in \cite{master-th} (see also \cite{Grain:2017dqa,Cable:2020dke}).
   
   \subsection{Brief overview} 
   
   A light field ($m^2 / H^2 \ll 1$) is split into a classical long wave length component $\overline{\Phi}$ and an operator  short distance contribution, 
   \begin{equation}
   \Phi = \overline{\Phi} (t, {\bf r}) + \int {d^3 {\bf k} \over (2\pi)^3 } \,  \theta (k - \epsilon a(t) H)     \,\left( a_{\bf k} \chi_k (t)  \, e^{i {\bf k} \cdot {\bf r}}+  a_{\bf k}^\dagger  \chi_k^* (t)  \, e^{-i {\bf k} \cdot {\bf r}}
   \right) \;,
   \end{equation}
   where $\epsilon \ll 1$,  $\theta (x)$ is the step function, and $a_{\bf k}, a_{\bf k}^\dagger $ are the annihilation and creation operators with the usual commutators.
   
   In the slow roll regime, $\ddot \Phi \ll H \dot \Phi$, and to leading order in $\epsilon$,  the equation of motion for the classical component is \cite{Starobinsky:1980te} 
   \begin{equation}
   \dot{\overline {\Phi}} =- {1\over 3H} V^\prime (\overline {\Phi} )   + f(t, {\bf r}) \;,
  \label{Langevin}
   \end{equation}
   with 
   \begin{equation}
   \langle      f(t, {\bf r})   f(t', {\bf r})  \rangle = {H^3\over 4\pi^2 } \delta (t - t')\;,
   \end{equation}
   which can be viewed as the ``noise'' term.
$\overline {\Phi}$ evolves according to the Langevin equation containing classical and stochastic forces. The probability of finding a specific value $\overline {\Phi}$ is described by the distribution function $\rho (\overline {\Phi},t)$, which is subject 
to the Fokker-Planck equation
\begin{equation}
{\partial \rho \over \partial t} = {H^3 \over 8 \pi^2}  \, {\partial^2  \rho \over \partial  \overline {\Phi}^2}  + {1\over 3 H} \, {\partial \over  \partial   \overline {\Phi} } \left(       V^\prime ( \overline {\Phi})\, \rho  \right) \;.
\label{Fokker}
\end{equation}
This distribution function describes the statistics of $ \overline {\Phi}$ over different Hubble patches.
   It then follows that \cite{Starobinsky:1980te} 
   \begin{equation}
   {d\over dt} \langle \overline {\Phi}^2 \rangle = -{2\over 3 } \,{m^2 \over H} \,  \langle \overline {\Phi}^2 \rangle  + {H^3 \over 4\pi^2 } \;.
   \end{equation}

   \subsection{Field size at the end of inflation}
   
   Solving for the time evolution of $\langle \overline {\Phi}^2 \rangle $ with proper boundary conditions at the beginning of inflation, one obtains  $ \langle \overline {\Phi}^2 \rangle $ at the end of inflation, which can eventually  be converted into 
   the particle density.
   For a massless field, 
   we get the ``random walk'' result 
    \begin{equation}
 \langle \overline {\Phi}^2 \rangle = \langle \overline {\Phi}^2 \rangle_0+  {H^3 \over 4\pi^2 } \, (t-t_0) \;,
 \label{masslessE}
   \end{equation}
   where $\langle \overline {\Phi}^2 \rangle_0$ is the initial value of $ \langle \overline {\Phi}^2 \rangle$ at $t=t_0$. 
   In the massive case, the solution is 
     \begin{equation}
 \langle \overline {\Phi}^2 \rangle =  {3 H^4 \over 8\pi^2 m^2} + \left(   \langle \overline {\Phi}^2 \rangle_0  -  {3 H^4 \over 8\pi^2 m^2}  \right)\, e^{    - {2m^2 \over 3H} \, (t-t_0)} \;.
  \label{massiveE}
   \end{equation}
   The solution {\it very slowly} tends to the asymptotic equilibrium value for $t-t_0 > H/m^2  $, 
    i.e.
     \begin{equation}
 \langle \overline {\Phi}^2 \rangle \rightarrow   {3 H^4 \over 8\pi^2 m^2}  ~~~{\rm for }~~~  \# {\rm ~of ~ e-folds} > {\cal O}\left( {H^2 \over m^2}  \right)\;,
   \end{equation}
   in agreement with (\ref{var-Phi}). 
   The required number of inflationary e-folds can easily be $10^{26}$, for example,  with $m=1$ GeV and $H=10^{13}$ GeV.
   On a shorter timescale, the size of the field is determined mostly by the  {\it pre-inflationary} initial condition at $t=0$,
     \begin{equation}
 \langle \overline {\Phi}^2 \rangle \simeq  \langle \overline {\Phi}^2 \rangle_0 \;. 
   \end{equation}
   Naturally, one does not expect it to be zero: after all, the initial value of the inflaton field is very large, probably beyond the Planck scale. In order not to affect inflation, the initial (and final) value of $\langle \overline {\Phi}^2 \rangle$
   is bounded by 
    \begin{equation}
V(\overline {\Phi})   \ll 3 H^2 M_{\rm Pl}^2 \;,
\label{Phi-bound}
   \end{equation}
   which only requires 
   \begin{equation}
 {\langle \overline {\Phi}^2 \rangle}   \ll   {H^2 M_{\rm Pl}^2 \over m^2} \;.
 \label{largePhi}
   \end{equation}
   The initial    long wavelength condensate evolves very slowly, due to a tiny classical force, and is almost unaffected by inflation, in analogy with the cosmological constant.  
Unless inflation proceeds  for a very (exponentially) long time, 
 one may expect a {\it much larger} field  at the end of inflation  compared to its equilibrium value, by a factor up to $M_{\rm Pl}/H$. 
  
  $\langle \overline {\Phi}^2 \rangle$ could also be far below $3 H^4 / (8\pi^2 m^2)$ if its initial value happens to be small. Expanding the exponential in (\ref{massiveE}) at ${2m^2 \over 3H} \, (t-t_0) \ll 1$,
  we recover the massless random walk result (\ref{masslessE}), to leading order in $m^2$. Since inflation requires at least 60 e-folds, $H(t-t_0) \gtrsim 60$, this sets the $lower$ bound on the field value,
  \begin{equation}
   {\langle \overline {\Phi}^2 \rangle}  > {\cal O} (H^2) \;.
   \label{lower-bound}
     \end{equation}
  The minimal value is achieved when $\langle \overline {\Phi}^2 \rangle_0 =  0$.
  
The fact that pre-inflationary initial conditions affect physical observables may  seem counterintuitive at first. 
However, this applies to the inflaton field as well. Although the  initial long-wavelength modes of the inflaton field do not affect the 
density perturbations, they determine the value of the scalar potential and thus the inflationary Hubble rate.  In the case of the spectator field, 
the condensate is not diluted by the expansion and  controls  the  eventual particle abundance.\footnote{The effect of the analogous  Higgs field condensate on reheating has been considered in \cite{Kaneta:2023kfv}.}

   The above considerations also apply to the effective mass generated during inflation via a non-minimal coupling to gravity $\xi$. 
  A small   $\xi$ is accounted for by replacing 
  \begin{equation}
   m^2 \rightarrow 12\, \xi\, H^2 
   \end{equation}
   in the above expressions, 
   as long as the physical mass satisfies $  m^2/H^2 \ll | \xi | \ll 1  $. A negative $\xi$ leads to a run-away behavior of   $ \overline {\Phi}$  in the limit of infinitely long inflation, 
   in agreement with the Bogolyubov approach of Sec.\,\ref{Bog-field}.
   However, 
   the result remains well defined for a finite duration of inflation, in which case $\xi <0$ simply increases the size of the condensate.

   \subsection{Particle density}
   
   The non-zero scalar field at the end of inflation carries energy which can subsequently be interpreted in terms of the particle number.
   Within a given Hubble patch, we may treat $\overline {\Phi}$ as a zero-mode field or a condensate. Its equation of motion with a zero initial velocity shows that
   $\dot{\overline {\Phi}}=0$ is the solution, given that $V^\prime (\overline {\Phi})$ can be neglected. Hence,
   after inflation ends, the condensate remains constant until the Hubble rate becomes comparable to the particle mass,
\begin{equation}
\overline {\Phi} \simeq {\rm const} ~~{\rm for }~~ {H>m}\;.
\end{equation}
When $H$ becomes comparable to $m$, the field starts oscillating in a quadratic potential. It can be interpreted as a collection of non-relativistic quanta with energy $m$, hence
\begin{equation}
n = {\rho \over m} = {1\over 2} m \overline {\Phi}^2 \;,
\end{equation} 
 where $\rho$ is the energy density of the scalar field. This happens when the scale factor is $a=a_m$ as in Eq.\,\ref{am}, after which the total particle number $a^3 n$ remains constant. 
 The value of $a_m$ depends on the scaling of the Hubble rate, i.e. whether the Universe is dominated by radiation or matter.
 
 \subsubsection{Radiation domination}
 
 In the radiation epoch,
 $H\propto 1/a^2$, so $a^3_m \,n(a_m) $ is given by
 \begin{equation}
 a^3 n = a_e^3 \, {H_e^{3/2} \over 2 m^{1/2}}\, \overline {\Phi}^2 \;.
 \end{equation}
 The result depends on the pre-inflationary initial condition for $ \overline {\Phi}$ as well as on the duration of inflation.
 In the extreme case of exponentially long inflation, we get 
 \begin{equation}
 a^3 n \rightarrow     {3\over 16 \pi^2 }  \,  a_e^3 \, {H_e^{11/2} \over  m^{5/2}}\,,
 \end{equation}
 in agreement with the Bogolyubov coefficient approach (\ref{a3-n-rad}) with $\kappa=1/2$.  This also applies to the case of a small positive $\xi$, while  for $\xi <0$ both approaches give  
 formally divergent results.

The above analysis assumes that the condensate is always  subdominant in the energy balance, otherwise it would trigger another inflation. Imposing  
${1\over 2} m^2 \overline {\Phi}^2  < 3 H^2 M_{\rm Pl}^2$
 at the condensate ``break-up''
point $H\sim m$, we get the consistency condition
\begin{equation}
  \overline {\Phi} <  {\cal O} (M_{\rm Pl }) \;,
 \end{equation}
 which is much stricter than (\ref{largePhi}).
The total particle number is bounded by
\begin{equation}
 a^3 n < a_e^3 \, {H_e^{3/2} \over  m^{1/2}}\,  {\cal O} ({\rm M_{Pl}^2})  \;.
 \end{equation}

 \subsubsection{Matter domination}

   The Hubble rate scaling is $H\propto 1/a^{3/2}$ and $H$ approaches $m$ during the matter domination epoch. Then
   \begin{equation}
 a^3 n = a_e^3 \, {H_e^{2} \over 2 m^{}}\, \overline {\Phi}^2 \;,
 \end{equation}
which, in the asymptotic equilibrium limit, gives
\begin{equation}
 a^3 n \rightarrow     {3\over 16 \pi^2 }  \,  a_e^3 \, {H_e^{6} \over  m^{3}}\,.
 \end{equation}
 This agrees with our previous result   (\ref{a3-n-mat}) for $\kappa=1/2$. As before, the non-minimal coupling case is  obtained by replacing two powers of the mass in the denominator:  $ m^2 \rightarrow 12 \xi H_e^2 $. 
 The consistency condition   $ \overline {\Phi} <  {\cal O} (M_{\rm Pl }) $ imposes an upper bound on the particle number, 
 \begin{equation}
 a^3 n <      a_e^3 \, {H_e^{2} \over  m^{}}\, {\cal O} ({\rm M_{Pl}^2}) \;.
 \end{equation}
 
 To summarize our findings, the Bogolyubov coefficient and Starobinsky stochastic approaches to gravitational particle production agree in the case of infinitely long inflation. In a more realistic situation, the Starobinsky  formalism is more convenient as it  
   readily includes non-trivial pre-inflationary initial conditions for the scalar field and accounts for a finite duration of inflation.

   \section{Constraints on dark relics}
   
   The abundance of stable particles produced by inflation cannot exceed that of dark matter. 
    The result depends both on the pre-inflationary initial conditions and duration of inflation, hence the best we can do is 
    $parametrize$ our ignorance in terms of the scalar 
     condensate  $ \overline {\Phi} $ at the end of inflation.
   
   The constraint is formulated in terms of the quantity $Y$, which is proportional to the total particle number,
   \begin{equation}
   Y = {n \over s_{\rm SM}} ~~, ~~ s_{\rm SM}={2\pi^2 g_*\over 45}\, T^3\;,
   \end{equation}
 where $s_{\rm SM}$ is the entropy density of the SM thermal bath and $g_*$ 
   is the effective number of degrees of freedom in the SM bath.
 $Y$ remains constant after reheating and is bounded by the dark matter abundance,
    \begin{equation}
   Y \leq  4.4 \times 10^{-10} \, {{\rm GeV} \over m}\;.
  \label{DMbound}
   \end{equation}
   
   The abundance calculation 
    involves the reheating temperature $T_R$.
   Reheating is defined as the point at which the energy density is transferred to the SM radiation and 
   the corresponding Hubble rate is 
    \begin{equation}
   H_R = \sqrt{\pi^2 g_* \over 90}\, {T_R^2 \over M_{\rm Pl}} \;.
  \label{HR}
   \end{equation}
   This can happen almost immediately after inflation or much later, if the inflaton couples very weakly to the SM fields. In what follows, we consider these two cases separately.

   \subsection{Radiation domination}
   
   Suppose that after inflation, the Hubble rate scales as $H \propto 1/a^2$. This can happen either due to fast reheating or the local inflaton potential being quartic, $V(\varphi) \propto \varphi^4$. The latter applies, for example, to Higgs inflation and alike,
   in which case the inflaton field oscillations in the quartic potential produce a radiation-like scaling of the energy density of the  Universe. 
      Since $Y$ remains constant after the condensate break-up, one can evaluate the abundance at temperature $T(a=a_m)$, which, by virtue of entropy conservation, is given by $T_R \, (a_R/a_m)$. 
 Taking $a_e =a_R$, we find 
 \begin{equation}
   Y \simeq 0.07 \times  {\overline {\Phi}^2\over m^{1/2} M_{\rm Pl}^{3/2} } \;,
   \label{Y-rad-dom}
   \end{equation}
   for $g_* \simeq 107$. We observe that the dark relic  abundance  is independent of $T_R$ and $H_e$.
   The same result is obtained if reheating occurs after the condensate starts oscillating, $a_R > a_m$, being preceeded by inflaton oscillations in the $\varphi^4$ potential.

    \subsection{Matter domination}
   
   This occurs when the inflaton oscillates in a local quadratic potential, $\varphi^2$, until it decays into SM radiation.
   Now the Hubble rate scales as $H \propto 1/a^{3/2}$ between $a_e$ and $a_R$. The condensate oscillations start before reheating, $a_m < a_R$, and the abundance can be computed at the reheating
   temperature $T_R$, noting that $a^3_m \, n (a_m) = a^3_R \, n(a_R)$.
   The result can be written as
 \begin{equation}
   Y \simeq 0.07 \times   {1\over \Delta} \,{ H_e^{1/2}\;  \overline {\Phi}^2\over m^{} M_{\rm Pl}^{3/2} } \;,
   \label{Y-matter-dom}
   \end{equation}
   with 
    \begin{equation}
  \Delta \equiv \sqrt{H_e\over H_R} \simeq {T_{\rm inst} \over T_R} \, \gg 1\;.
  \label{Delta}
   \end{equation}
Here $T_{\rm inst}$ denotes the instant reheating temperature, which  corresponds  to the (hypothetical) instant transfer of the inflaton energy at the end of inflation to the SM radiation.
   
   We observe that the relic abundance is diluted in the case of late reheating, $\Delta \gg 1$. This factor can be as large as $10^{17}$ for MeV-scale reheating temperatures. Such late reheating
  occurs if a non-relativistic inflaton oscillates undisturbed in the  $\varphi^2$ potential, until it decays due to a   very small  coupling to the SM fields, primarily the Higgs.
  The decay takes place when the Hubble rate approaches the inflaton decay width,
$
 H \sim \Gamma (\varphi \rightarrow hh) \;,
 $
which can be made  arbitrarily small by reducing the Higgs inflaton coupling $\varphi H^\dagger H$.

   \subsection{Implications}
   
   We observe that the dark relic abundance is quadratically sensitive to the unknown post-inflationary condensate $ \overline {\Phi}$, which is only bounded by the inflationary considerations and consistency,
   \begin{equation}
    H_e \lesssim  \overline {\Phi} \lesssim  M_{\rm Pl} \;.
   \end{equation}
   Hence,  the $Y$ predictions vary by at least 10 orders of magnitude.
   
   The dark relic abundance in the radiation  dominated and matter dominated Universes exhibit different features. In particular, the radiation case is very ``stiff'' being independent of the reheating temperature.
   Let us consider some practical implications of our results.  
 
     {\bf  \underline{Radiation domination}.} Using the dark matter bound (\ref{DMbound}), we get the following constraint on the condensate size
    \begin{equation}
  \overline {\Phi} <  {5\times 10^9 \over (m/ {\rm GeV})^{1/4}}\; {\rm GeV}\;.
  \label{Phi-free-rad}
   \end{equation}
   This is incompatible with large field inflation $H_e\sim 10^{13}-10^{14}$ GeV, unless the dark relic is exceptionally light, 
   $$m \ll \; {\rm eV}\,.$$ 
   Indeed, the lower bound (\ref{lower-bound}) requires $ \overline {\Phi} > H_e$, while the mass dependence  in (\ref{Phi-free-rad})  is very mild.
   An even stronger bound is obtained if one assumes the asymptotic value $    \overline {\Phi}^2    =    {3 H_e^4 \over 8\pi^2 m^2} $.
   Indeed, combining the constraint on $Y$ with $m\ll H_e$, one finds a mass-independent bound $H_e\lesssim 10^8\,$GeV.
    This conclusion is independent of the reheating details, as long as the Universe energy density exhibits radiation-like scaling.

    {\bf  \underline{Matter domination}.} In this case, the constraint on the condensate size is $m$-independent,
     \begin{equation}
  \overline {\Phi} <  \Delta^{1/2} \times {5\times 10^9 \over (H_e/ {\rm GeV})^{1/4}}\; {\rm GeV}\;.
   \label{matter-Phi}
   \end{equation}
   The lower bound $ \overline {\Phi} > H_e$ then implies
   \begin{equation}
   H_e \lesssim \Delta^{2/5} \times 5 \cdot 10^7\; {\rm GeV}\;.
      \end{equation}
     Therefore, large field inflation with    $H_e\sim 10^{13}-10^{14}$ GeV is only possible if 
       \begin{equation}
   \Delta \gtrsim 10^{15} \;,
      \end{equation}
   assuming that a stable dark relic with $m\ll H$ exists. 
   Its abundance must be  diluted by a very large factor $\Delta $ in order to be consistent with observations. Since the maximal instant reheating temperature is of order $10^{15}\;$GeV,
   such dilution implies 
    a {\it very low} reheating temperature in the GeV range or below, as seen from (\ref{Delta}).
    
   It is important to emphasize that these  constraints are $optimistic $ and assume the lowest possible value of $\overline {\Phi} $ at the end of inflation. In reality, one could expect much larger values, up to 
   $\overline {\Phi}  \sim M_{\rm Pl}$ without affecting the inflationary predictions.
   The bound (\ref{matter-Phi}) can also be written as
    \begin{equation}
  \overline {\Phi} \lesssim      { 10^{14} \over (T_R/ {\rm GeV})^{1/2}}\; {\rm GeV}\;.
   \end{equation}
Then, a Planckian value for $  \overline {\Phi}$ would require a tiny reheating temperature, far below the BBN bound. This would make the existence of stable dark relics with $m \ll H$ incompatible 
with inflation and standard cosmology.

  In the above considerations,  $m$ represents the physical mass, while  the non-minimal coupling to gravity  $\xi $   does not appear explicitly as it only affects  $  \overline {\Phi}$ in our approximation.

   \subsection{Achieving a low reheating temperature}
   
   The reheating temperature is determined by the inflaton decay rate into the SM states and  only required to be above 4 MeV by observations \cite{Hannestad:2004px}. Since the inflaton is expected to be a singlet under the SM symmetries, the $only$ renormalizable couplings between the inflaton $\varphi$
   and the SM fields are \cite{Lebedev:2021xey}
   \begin{equation}
   V_{\varphi h}= \sigma_{\varphi h} H^\dagger H \varphi + {1\over 2} \lambda_{\varphi h} H^\dagger H \varphi^2 \;.
   \end{equation}
   The second term leads to early time Higgs production, which can be quite intensive, depending on the coupling, and potentially lead to a quasi-equilibrium state of the $\varphi - h$ system. However, as the Universe expands, the heavy inflaton comes to dominate again and the energy density of the relativistic Higgs quanta 
 red-shifts away. Hence, the main driver of  reheating is the trilinear coupling $\sigma_{\varphi h} $, which leads to late-time decay of the inflaton quanta.
   
   As long as the inflaton is much heavier than the Higgs,  $m_\varphi \gg m_h$, the perturbative decay width is given by 
    \begin{equation}
  \Gamma (\varphi \rightarrow h_i h_i) = {\sigma_{\varphi h}^2 \over 8\pi m_\varphi } \;,
     \end{equation}
for 4 Higgs degrees of freedom at high energies.
When the Hubble rate approaches the decay width, $H_R \sim \Gamma (\varphi \rightarrow h_i h_i)$,  reheating occurs with the 
reheating temperature given by (\ref{HR}). This implies
 \begin{equation}
  T_R \sim 2\times 10^8\; \sigma_{\varphi h}  \; \sqrt{{\rm GeV} \over m_\varphi} \,.
     \end{equation}
Therefore, for a heavy inflaton, $m_\varphi \sim 10^{13}\;$GeV, one has $ T_R \sim 10^2 \, \sigma_{\varphi h} $. A GeV reheating temperature would require $\sigma_{\varphi h} $ in the MeV range.
While the coupling appears very small, it is radiatively stable and can be justified by approximate $Z_2$ symmetry under which
 \begin{equation}
  \varphi \rightarrow -\varphi\;.
     \end{equation}
Therefore, a very low $T_R$ can be achieved in a straightforward  manner.

Once the relics produced by inflation are diluted via a low reheating temperature, dark matter can be regenerated through the ``stronger coupling''  freeze-in mechanism \cite{Cosme:2023xpa}, 
which can operate even at very low temperatures  $\gtrsim {\cal O}({\rm MeV})$ \cite{Lebedev:2024vor}.

       \section{Weakly coupled scalars: Starobinsky-Yokoyama approach}
   
   So far we have considered production of scalars whose couplings  can be neglected. 
   Let us now include a   weak but non-negligible scalar self-coupling \cite{Starobinsky:1994bd},
   \begin{equation}
   V(\Phi) = {m^2\over 2}   \Phi^2 + {\lambda\over 4}\Phi^4\;,
   \end{equation}
   with $\lambda \ll 1$.
   
   For small $\lambda$, the Langevin (\ref{Langevin})
and Fokker-Planck ({\ref{Fokker}}) equations still apply. Multiplying (\ref{Fokker}) by $  \overline {\Phi}^2 $ and integrating over     $ \overline {\Phi}$ from $-\infty$ to $\infty$, we obtain 
the evolution equation for $\langle \overline {\Phi}^2 \rangle$. In order to estimate the effect of the self-coupling, let us resort to the {\it  Hartree-Fock } or {\it  Gaussian} approximation
$\langle \overline {\Phi}^4 \rangle=     3 \langle \overline {\Phi}^2 \rangle^2$, in which case \cite{Starobinsky:1994bd}
\begin{equation}
   {d\over dt} \langle \overline {\Phi}^2 \rangle = -{2\over 3 } \,{m^2 \over H} \,  \langle \overline {\Phi}^2 \rangle   -      {2 \lambda    \over H} \,  \langle \overline {\Phi}^2 \rangle^2+        {H^3 \over 4\pi^2 } \;.
   \end{equation}
   This is to supplemented with the initial {\it pre-inflationary} condition   $ \langle \overline {\Phi}^2 \rangle (t=0) = \langle \overline {\Phi}^2 \rangle_0$.
  The equilibrium state is obtained by setting the right hand side to zero, which implies that 
  the asymptotic behavior of $  \langle \overline {\Phi}^2 \rangle $ is
\begin{equation}
   \langle \overline {\Phi}^2 \rangle   \rightarrow {H^2 \over \sqrt{8} \pi}\, {1\over \sqrt{\lambda}} \;,
  \label{Phi-lambda}
   \end{equation}
as long as the field is light enough, $m^4/H^4 \ll \lambda$. This corresponds to the field obtaining an effective mass squared $m^2_{\rm eff} = {3H^2 \over \sqrt{8} \pi } \, \sqrt{\lambda}$.
The equilibrium state is approached on a  characteristic timescale  $(\sqrt{\lambda} H)^{-1}   \gg H^{-1}$, which is much longer than the Hubble time  for $\lambda \ll 1$.
   Therefore, as before, the size of the condensate is determined primarily by its initial value  $ \langle \overline {\Phi}^2 \rangle_0$, unless inflation is very long. Its minimal value is of order $H^2$ at small times. 
   
   A more careful analysis of the Fokker-Planck equation shows that the equilibrium distribution is non-Gaussian \cite{Starobinsky:1994bd}, 
    \begin{equation}
  \rho   ( \overline {\Phi}) \propto \exp \left({-  {4 \pi^2 V(\overline {\Phi} ) \over 3 H^4}} \right)\;,
   \end{equation}
   which means that $  \langle \overline {\Phi}^2 \rangle  \simeq 0.132 {H^2 \over \sqrt{\lambda}}$, not far from the above simple estimate. 
   In any case, the actual condensate value at the end of inflation remains unknown and, as before, is only constrained by
    \begin{equation}
    H_e \lesssim  \overline {\Phi} \lesssim  M_{\rm Pl} \;.
    \label{ineq-y}
   \end{equation}
The lower bound is imposed by inflationary dynamics at $\lambda \ll 1$, while the upper bound is required by consistency (see below).

   \subsection{Dark relic abundance}
   
   The dark relic abundance in   the Starobinsky-Yokoyama approach was studied in \cite{Peebles:1999fz,Markkanen:2018gcw} and, in a more general setting,  \cite{Lebedev:2022cic}.
   
   Consider the case of matter dominated Universe after inflation, $H\propto a^{-3/2}$.  In the Hartree-Fock approximation, the scalar self coupling induces an effective mass, 
   \begin{equation}
 m^2_{\rm eff} = m^2 + 3 \lambda  \overline {\Phi}^2 \;,
  \end{equation} 
  with $\lambda  \overline {\Phi}^2 \gg m^2$.
   We assume the field to be light, $m_{\rm eff}  \ll H_e$, and therefore  require 
   \begin{equation}
 \lambda  \overline {\Phi}^2  \ll H_e^2\;.
 \label{bound-lambda}
  \end{equation} 
   This condition  is, for example, trivially satisfied for the asymptotic equilibrium value   (\ref{Phi-lambda}). It also ensures that the spectator  does not affect inflationary dynamics as long as $\overline {\Phi} \lesssim  M_{\rm Pl}$.
   
   Since the field is light, $ \overline {\Phi}$ remains frozen for some time  after inflation. When the Hubble rate decreases to the level of $m_{\rm eff}  $, the average field starts oscillating in the quartic potential since 
   $\lambda  \overline {\Phi}^2 \gg m^2$.      We denote the corresponding scale factor $a_{osc}$ such that
   \begin{equation}
   H_{osc} \sim m_{\rm eff} 
   \end{equation}
   at this stage.      
   The condensate remains a subdominant energy component if $3 H_{osc}^2 M_{\rm Pl}^2 >  {\lambda\over 4} \overline \Phi^4$, which requires $\overline {\Phi} \lesssim  M_{\rm Pl} $, as stated in (\ref{ineq-y}).
       From this point on,  the field  amplitude decreases as $1/a$.
     At a later  stage, the quadratic and quartic terms in the scalar potential become comparable,
      \begin{equation}
   {m^2\over 2}  \overline \Phi^2  \sim  {\lambda\over 4} \overline \Phi^4\;.
   \end{equation}
  This happens at $a=a_m$, after which the potential is dominated by the quadratic term. Therefore, the field becomes effectively a collection of non-relativistic particles with the particle density $n\simeq  V( \overline \Phi)/m$.
  Subsequently, reheating occurs at $a=a_R>a_m$. The final result remains the same if one assumes $a_R<a_m$.
   
   Thus, we have the following stages in the system evolution: 
    \begin{equation}
a_e \rightarrow a_{osc} \rightarrow a_m \rightarrow a_R~~,
\end{equation}
with the field amplitude scaling
     \begin{equation}
 \overline \Phi  ~{\stackrel{ a^0} { \longrightarrow   }  }   ~ \overline \Phi_{ osc}    ~
 \stackrel{  a^{-1}} \longrightarrow    \overline \Phi_{m} ~{\stackrel{ a^{-3/2}} { \longrightarrow   }  }    ~  \overline \Phi_{R} ~~.
\end{equation}
Computing the particle abundance at reheating, one finds in the case of {\it matter domination},
\begin{equation}
   Y \simeq 10^{-2} \times   {1\over \Delta} \,{ H_e^{1/2}\;  \overline {\Phi} \over \sqrt{\lambda} \,M_{\rm Pl}^{3/2} } \;.
   \end{equation}
   Interestingly, this expression has the same form and also is numerically close to our free-field result (\ref{Y-matter-dom}) with $m \sim \sqrt{\lambda} \,\overline {\Phi}$.

  Since  $  H_e \lesssim  \overline {\Phi}$,
  the bound on the dark matter abundance   in the matter-dominated case    requires \cite{Lebedev:2022cic}
     \begin{equation}
m \, \lambda^{-1/2} \lesssim {\rm few} \times10^{-8}\times \Delta_{\rm } \,\left( {M_{\rm Pl}\over H_{\rm end}}\right)^{3/2} 
  \,{\rm GeV} \;.
 \label{SY-mat}
 \end{equation}
   High scale inflation then implies $m \, \lambda^{-1/2} \lesssim  \Delta  \,{\rm GeV}$.
   Therefore, heavy and/or feebly coupled dark relics are only allowed if the dilution factor is very large $\Delta \gg 1$, implying a low $T_R$.
   The origin of the inverse dependence on the size of the coupling can be traced to the   particle density: $n(a_m)= m^3/\lambda$, such that more particles are produced at weaker couplings. Note that $\lambda $ cannot be arbitrarily small:
   we require $\lambda  \overline {\Phi}^2 \gg m^2$ in our analysis. 
   The upper bound on $\lambda$ is imposed  by   (\ref{bound-lambda}), i.e. $\lambda \ll H_e^2 / \overline {\Phi}^2$, as well as by non-thermalization of $\Phi$.
 Indeed, a significant self-coupling  would thermalize the relic invalidating our estimates based on a conserved particle number.
   The corresponding bounds on the coupling are presented in \cite{Arcadi:2019oxh}, e.g. for $m\sim 1\;$GeV, the non-thermalization constraint requires $\lambda < 10^{-3}$.

   In the {\it radiation domination} case, $H \propto a^{-2}$ and $a_e=a_R$, and   the constraint is stronger since there is no dilution factor.
   We find
   \begin{equation}
   Y \simeq 10^{-2} \times    { \overline {\Phi}^{3/2} \over \lambda^{1/4} \,M_{\rm Pl}^{3/2} } \;.
   \end{equation}
   Again, the abundance has the same form and is also numerically close to our free scalar result (\ref{Y-rad-dom}) with $m \sim \sqrt{\lambda} \,\overline {\Phi}$.
   Applying the lower bound $  H_e \lesssim  \overline {\Phi}$, the consequent constraint on the mass-coupling combination is
     \begin{equation}
m \, \lambda^{-1/4} \lesssim  10^{-8}\times  \left( {M_{\rm Pl}\over H_{\rm end}}\right)^{3/2} 
  \,{\rm GeV} \;.
  \label{SY-rad}
   \end{equation}
 High scale inflation then requires 
   $m \, \lambda^{-1/4} \lesssim  10^{-1} \,{\rm GeV}$, meaning that stable, feebly coupled relics can have at most MeV scale masses.

   \section{Inflation-induced   mass: 
   suppressing inflationary  and enhancing post-inflationary  particle production}
   
   We have so far considered production of light particles, $m\ll H$, during inflation. These results do not apply if the dark scalar $\Phi$ attains  a large inflation-induced mass above the Hubble scale, $m_{\rm eff}\gtrsim H$,
   suppressing particle production.
   This can happen, for example, due to its positive coupling to the inflaton $\varphi$, 
    \begin{equation}
   V_{\Phi \varphi}= {1\over 4} \lambda_{\Phi \varphi}\, \varphi^2 \Phi^2 \;,
   \end{equation}
   or a significant non-minimal coupling to gravity $\xi \gtrsim {\cal O}(1)$. 
   Both of these couplings lead, however, to efficient  particle production immediately after inflation, i.e. during the inflaton oscillation epoch.
   The $\xi$-induced production is very efficient \cite{Bassett:1997az} since $\xi \Phi^2 R $ generates a tachyonic mass term when the sign of $R$ alternates.
   On the other hand,  $ \lambda_{\Phi \varphi} >0$  always leads to a positive mass term,  making the effect milder.
   
    To be conservative, let us focus on postinflationary particle production induced by $ \lambda_{\Phi \varphi} $. 
 Its  efficiency  depends on the relation between the induced scalar mass $m_{\rm eff}  \sim \sqrt{\lambda_{\Phi \varphi}}\, |\varphi |$ and the inflaton mass $m_\varphi$, which can be either bare or
 effective $m_\varphi \sim \sqrt{\lambda_{\varphi}}\, |\varphi |$ in the case of the quartic local potential. If 
 $m_{\rm eff} \gg m_\varphi$ after the end of inflation, particle production is very efficient being enhaced by collective effects due to resonances
 \cite{Kofman:1997yn,Greene:1997fu}. At yet larger couplings,  the inflaton-dark scalar system
 can reach quasi-equilibrium where the energy is distributed equally among all the degrees of freedom \cite{Lebedev:2021tas}. On the other hand, if 
 $  m_{\rm eff} \lesssim  m_\varphi $  
 shortly after inflation, production of the $\Phi$-quanta is slow and can be treated perturbatively.

 \subsection{Weak coupling}
 
   Consider the small   $\lambda_{\Phi \varphi}$   coupling regime in which the perturbative approach is adequate \cite{Dolgov:1989us,Traschen:1990sw,Ichikawa:2008ne}.
   After inflation, the inflaton field undergoes oscillations around the minimum of the potential.
   It can be expanded as 
    \begin{equation}
\varphi^2(t) = \sum_{n=-\infty}^\infty \zeta_n e^{-in \omega t} \;,
 \end{equation}
 where
 the frequency $\omega$ is determined by the (effective) inflaton mass and 
  the coefficients $\zeta_n$ are slow functions of time: they scale as $1/a^2$ or $1/a^{3}$ depending on whether the background
 is radiation- or matter-dominated.
  A time dependent background naturally leads to particle production and if $\Phi$ is lighter than the inflaton, it will be pair-produced.
  
 As stated earlier, we focus on the regime where the dark scalar becomes lighter than the inflaton shortly after inflation,  $ m_{\rm eff} \lesssim  m_\varphi $, such that 
  the above reaction is allowed kinematically. 
 Creation of a two--particle  state with momenta {} $p,q$ from the vacuum is described by the amplitude    \cite{Peskin:1995ev}
   \begin{equation}
-i \int_{-\infty}^\infty dt \langle f | V_{\Phi \varphi}(t) | i \rangle = - i \,{ \lambda_{ \Phi \varphi} \over 2 } \, (2\pi)^4 \delta ({\bf{p}} + {\bf{q}})  \sum_{n=1}^\infty \zeta_n \delta(E_p +E_q -n \omega) \;.
 \end{equation}
  The resulting reaction rate  for $\Phi$-pair production per unit volume is 
    \begin{equation}
 \Gamma =    \sum_{n=1}^\infty  {1\over 2} \int |{\cal M}_n|^2 d \Pi_n =
 {\lambda_{ \Phi \varphi}^2 \over 64 \pi  } \sum_{n=1}^\infty  |\zeta_n|^2 \sqrt{1- \left(   {2m \over n \omega} \right)^2} \; \theta(n\omega -2 m) \;,
 \label{Gamma-ss}
  \end{equation}
where ${\cal M}_n$ is the invariant amplitude for the $n$-th inflaton mode and $\Pi_n$ represents the corresponding phase space.
To account for adiabatic Universe expansion, the coefficients $\zeta_n$ 
get rescaled as
\begin{equation}
\zeta_n \rightarrow \zeta_n/a^l \;,
\end{equation}
with $l= 3$ and $2$ for the matter- and radiation-dominated Universe, respectively. 
   
  The reaction rate  depends on the $local$  inflaton  potential, which can be quadratic, $V(\varphi) = m^2_\varphi \varphi^2/2$, or quartic,  $V(\varphi) = \lambda_\varphi \varphi^4/4$.
   In the former case,      $\varphi(t) \simeq  \varphi_0\,\cos m_\phi t $,  
   and the Universe is matter-dominated.
   There is just one harmonic that contributes to the reaction, $|\zeta_1|= \varphi_0^2/4$ and $\omega = 2m_\varphi$.
   Neglecting the final state mass and integrating the Boltzmann equation 
   \begin{equation}
   \dot n + 3Hn = 2 \Gamma \;,
   \end{equation}
   one finds, in the case of matter domination \cite{Lebedev:2022cic},
     \begin{equation}
   Y\simeq 2 \cdot 10^{-4} \times {1\over \Delta }\; {\lambda_{\Phi \varphi}^2 \, \varphi_0^4 \over H_e^{5/2} M_{\rm Pl}^{3/2} } \;. 
     \end{equation}
Since the production  rate drops fast with $a$, namely $a^{-6}$,   the result is dominated by the {\it early time} contribution immediately after inflation. 

For the quartic local inflaton potential, the Universe is effectively radiation-dominated. The inflaton oscillates according to the Jacobi cosine function, which one can approximate by the first harmonic
with $|\zeta_1|\simeq  0.14\, \varphi_0^2$.
The resulting dark scalar abundance is \cite{Lebedev:2022cic}
    \begin{equation}
   Y\simeq   10^{-4} \times { \Delta }\; {\lambda_{\Phi \varphi}^2 \, \varphi_0^4 \over H_e^{5/2} M_{\rm Pl}^{3/2} } \;. 
   \label{Y-rad-2}
     \end{equation}
An interesting feature in this case is that the factor $\Delta $ appears in the numerator and increases $Y$. The reason is that the total particle number $n a^3$ grows linearly with $a$ such that the result
is dominated by late times. Particle production stops either due to inflaton decay or loss of coherence of the inflaton background. The latter is due to inflaton self-interaction  which induces fragmentation
and breakdown of coherent oscillations. The duration of coherent oscillations depends logarithmically on $\lambda_\varphi $ such that the 
 corresponding Hubble rate $H_*$ and the $\Delta$-factor  satisfy \cite{Khlebnikov:1996mc}
 \begin{equation}
\Delta =\sqrt{H_e \over H_*} \sim z_* ~~, ~~ z_* \simeq 76- 14.3 \ln \lambda_\varphi \;,
 \end{equation}
for $\varphi_0$ not far from the Planck scale. In practice, one expects $\Delta \sim {\cal O}(10^2)$. 
However, if the inflaton decays in the SM states faster  than it loses coherence due to self-interaction, then $\Delta$ in (\ref{Y-rad-2}) 
is given by  the usual expression $\sqrt{H_e/H_R}$, as before.
   
   The consequent constraint on the inflaton coupling to $\Phi$ can be put in a universal form,
   \begin{equation}
    \lambda_{\Phi \varphi} \lesssim 10^{-3} \times \Delta^{\pm 1/2} \, {   H_e^{5/4} M_{\rm Pl}^{3/4}  \over \varphi_0^2}\; \sqrt{{\rm GeV} \over m} \;,
     \end{equation}
 with ``+'' for matter domination and ``-'' for radiation domination.

   To suppress particle production during inflation, we require
   $$ m_{\rm eff} \gtrsim H\;, $$
   such that $\Phi$ behaves as a classical field locked at the origin. This implies, in particular, 
   $ \lambda_{\Phi \varphi}  \gtrsim 2H_e^2 / \varphi_0^2$. Since the inflaton potential at large field values is concave, this inequality guarantees that
   $\Phi$ remains heavy throughout inflation.  Combining the upper and lower limits on $ \lambda_{\Phi \varphi} $, one obtains a consistency condition
   $   10^{-3}\times \Delta^{\pm 1/2} \left(   {M_{\rm Pl} \over H_e}   \right)^{3/4}   \sqrt{{\rm GeV} \over m} \gtrsim 1   $. 
   For standard high scale inflation, this condition 
  requires a nontrivial dilution factor 
            for any scalar mass above a GeV,    $\Delta \sim m/{\rm GeV}$,    assuming matter-dominated expansion. In the radiation-dominated case, 
            only sub-GeV stable particles are allowed.
 Taking $\varphi_0 \sim M_{\rm Pl}$ and $H_e \sim 5\times 10^{13}\,$GeV as the benchmark values, one obtains the following 
  bounds on the coupling:
    \begin{equation}
  10^{-9}  \lesssim \lambda_{\Phi \varphi} \lesssim 10^{-9} \times \Delta^{\pm 1/2} \,   \sqrt{{\rm GeV} \over m} \;.
     \end{equation}
   We thus conclude that, given a long enough matter-dominated period after inflation, suppression of particle production is possible within a limited range of the inflaton-dark scalar couplings, depending on the dilution factor.
   These perturbative considerations only apply if $m_{\rm eff} \lesssim m_\varphi$, which imposes a further upper bound on  the coupling:
  $ \lambda_{\Phi \varphi}  \lesssim (m_\varphi / \varphi_0)^2  $ for the local quadratic inflaton potential\footnote{If the condition  $m_{\rm eff}\lesssim m_\varphi$ is not satisfied initially, it will be as the amplitude of the inflaton oscillations decreases.}           
   and $ \lambda_{\Phi \varphi}  \lesssim  \lambda_{\varphi} $ for the quartic one.

  \subsection{Stronger coupling}
  
  For  larger $ \lambda_{\Phi \varphi}$,  
  $$    m_{\rm eff} \gg H\,, \, m_\varphi   \;, $$ 
  during and shortly after inflation, 
whereas
  the  simple perturbative approach now breaks down. Lattice simulations show that  $ \lambda_{\Phi \varphi} \gtrsim 10^{-6}$, for typical parameter values, leads to explosive post-inflationary particle production
  which backreacts on the inflaton background and results in 
  quasi-thermalization of the system. That is, shortly after inflation, both the inflaton and the dark scalar become relativistic and share the energy density in approximately equal proportions. The resulting abundance of $\Phi$ becomes {\it coupling-independent}
   \cite{Lebedev:2021tas}.
 
  This can be see as follows.
   During the quasi-equilibrium stage, the number densities of the $\varphi$ and $\Phi$ quanta are similar, $n_\varphi \sim n$. 
 As lattice simulations show, 
    the particle number is approximately conserved thereafter,      
    such that the above  relation persists until reheating, i.e. inflaton decay.
   Since the inflaton is heavier, it becomes non-relativistic at $a=a_*$ and starts dominating the energy density of the Universe from this point on.
   The reheating temperature is then determined by the energy density stored in these non-relativistic quanta, $T_R^4 \propto n_\varphi (a_*) \, m_\varphi \, (a_*/a_R)^3$. The inflaton number density at $a=a_*$ determines the Hubble rate:
   $n_\varphi (a_*) \, m_\varphi \sim 3 H_*^2 M_{\rm Pl}^2$. Solving for $n_\varphi (a_*) \sim n(a_*)$ and using the scaling $H \propto a^{-3/2}$ between $a_*$ and $a_R$, one finds a coupling-independent abundance
   \cite{Lebedev:2021tas}, 
   \begin{equation}
Y \sim {0.4\over \Delta} \; {M_{\rm Pl}^{1/2} H_e^{1/2} \over m_\varphi} \;,
\end{equation}
with the usual $\Delta \equiv \sqrt{H_e/H_R}$.

This result applies to both $\varphi^2$ and $\varphi^4$ local inflaton  potentials, as long as the inflaton is heavier than the dark relic.
  The consequent constraint on the dilution factor is very strong. For the typical parameter values $H_e \sim 5\times 10^{13}\,$GeV, $m_\varphi \sim 10^{13}\,$GeV, we have 
  \begin{equation}
\Delta \gtrsim 10^{12}\; {m\over {\rm GeV}}\;,
\end{equation}
requiring  a low reheating temperature, $T_R <\,$TeV, 
if there exist stable scalars 
  at or above the GeV scale. For a lighter inflaton, as in the $\varphi^4$  case, the constraint is yet much stronger.

  Similar results hold for the effective mass term generated by the non-minimal coupling to gravity $\xi$. During inflation, it induces $m_{\rm eff}^2 = 12 \xi H^2$, so $\xi> 10^{-1}$ makes the field ``heavy''.
  After inflation, one can expand the Lagrangian in small $\varphi < M_{\rm Pl}$, which  generates an effective inflaton-dark scalar coupling in the Einstein frame. In particular, for a local quadratic inflaton potential, one has $ \lambda_{\Phi \varphi} \sim \xi m_\varphi^2/M_{\rm Pl}^2$. In addition to that, it produces a derivative coupling of the form $-\xi \dot \varphi^2 \Phi^2 $ \cite{Ema:2017loe}, which makes particle production more efficient compared to that in the pure $ \lambda_{\Phi \varphi} $-coupling case. Similarly, the stronger coupling regime $\xi \gtrsim {\cal O}(100)$
  leads to quasi-equilibrium in the $\varphi-\Phi$
 system \cite{Lebedev:2022vwf}, resulting in large dark relic abundance.  Thus, many of our conclusions apply to the   $\xi$-induced effective mass as well.

To summarize, although the inflationary fluctuations and particle production get  suppressed by an induced mass term, the preheating dynamics reintroduces the problem. 

\subsection{Quantum gravity induced operators}

In addition to the mechanisms  discussed above, particles can be copiously produced via higher dimensional operators generated by classical and quantum gravitational effects. 
Quantum gravity is believed to lead to all couplings consistent with gauge symmetry and therefore expected to induce the inflaton interaction with  the dark scalar.
 Although the structure and the size of such interaction is unknown, one may resort to the effective field theory expansion in order to analyze its effect.
 
After inflation, the inflaton field amplitude decreases and,  in the regime 
$\varphi < M_{\rm Pl}$,  the Lagrangian can be expanded in powers of $\varphi /M_{\rm Pl}$.  Among others, one expects interactions of the form \cite{Lebedev:2022ljz,Lebedev:2022cic},  
\begin{equation}
{\varphi^4 \Phi^2 \over M_{\rm Pl}^2} ~,~ {\varphi^6 \Phi^2 \over M_{\rm Pl}^4}~,~ {\varphi^8 \Phi^2 \over M_{\rm Pl}^6}~,~...
\label{op-list}
\end{equation}
These are very efficient in particle production. Indeed, as long as $\varphi$ is not too far below the Planck scale, such operators behave similarly to the coupling $\varphi^2 \Phi^2$
considered above. Denoting the Wilson coefficient of the operator  $\varphi^4 \Phi^2/M_{\rm Pl}^2$ as ${\cal C}$, at weak coupling one finds \cite{Lebedev:2022cic}
\begin{equation}
|{\cal C}| \lesssim 10^{-3}\, \Delta^{1/2} {H_e^{5/4} M_{\rm Pl}^{11/4} \over \varphi_0^4}\, \sqrt{{\rm GeV}\over m} \;,
\end{equation}
where $\Delta =1$ for radiation domination and $\Delta \equiv \sqrt{H_e/H_R}$ for matter domination.
For the high scale inflation typical parameter  values, 
this implies $|{\cal C}| \lesssim 10^{-8}\, \Delta^{1/2} \sqrt{{\rm GeV}/ m}$.
The constraint is {\it very strong} requiring the Wilson coefficient to be tiny unless 
  $\Delta \sim 10^{15}$ or the dark scalar is extremely light. Even though the operator is Planck-suppressed, its effect on particle production is powerful.
  A similar conclusion applies to other operators in (\ref{op-list}), while derivative    $\varphi-\Phi$ couplings have a much milder effect.  Interestingly,
    the interaction $\varphi^4 \Phi^2/M_{\rm Pl}^2$     with the Wilson coefficient $|{\cal C}| \gtrsim 10^{-4}$   can bring the inflaton-dark scalar system into a quasi-equilibrium state,
    in which case the relic abundance becomes independent of ${\cal C}$  \cite{Lebedev:2022ljz}.

We note that the dark relic production at this stage can be viewed as gravity-mediated inflaton annihilation \cite{Kaneta:2022gug,Garcia:2022vwm}, including the narrow resonance regime \cite{Dvali:2022vzz}.

The Planck-suppressed operators are also efficient in producing fermions $\Psi$ \cite{Koutroulis:2023fgp}. Although the inflationary fermion production is suppressed by the fermion mass \cite{Chung:2011ck}, postinflationary dynamics lead to efficient $\Psi$ production via operators of the type 
$\varphi^2 \bar \Psi \Psi /M_{\rm Pl}$. These can generate all of the required dark matter even for small values of the Wilson coefficients \cite{Koutroulis:2023fgp}.
We emphasize that quantum gravity $violates$ conformal invariance such that the couplings are not subject to the corresponding constraints.

\subsection{Dark matter}

A special case of a dark relic is dark matter, whose abundance is
 \begin{equation}
   Y =  4.4 \times 10^{-10} \, {{\rm GeV} \over m}\;.
   \end{equation}
As is clear from the above considerations, such a value can be obtained for a sufficiently low $T_R$. 
An additional constraint is imposed by the isocurvature perturbation bound, see e.g. \cite{Garcia:2025rut} for a recent analysis.
Generally, it is difficult to circumvent this bound if dark matter is generated by the de Sitter fluctuations since these are not correlated with the inflaton fluctuations. On the other hand, the above preheating 
dynamics can readily be responsible for consistent dark matter production because it is determined by the inflaton field.

\section{Summary of results}
 
Below we list the main results of our work. We focus on inflationary production of light ($m\ll H$), free or feebly interacting  scalar fields in the high scale inflation framework. 
The scalar is  also allowed to have a small non-minimal coupling to gravity, $|\xi| \ll 1$, away from the conformal point.\footnote{At the conformal point $\xi=1/6$, particle production is suppressed. However,  the existence of the Planck scale shows that quantum gravity 
  violates 
conformal invariance strongly. Hence, the choice $\xi\sim1/6$ does not appear well motivated in a realistic setting.} 
We find that:

 \begin{itemize}
  \item{the Bogolyubov coefficient and Starobinsky approaches to inflationary particle production agree in the limit of infinitely long inflation. The standard Bogolyubov coefficient  approach assumes the Bunch-Davies vacuum at the beginning of inflation,
  which corresponds to the infinite past. The Starobinsky stochastic approach, on the other hand, naturally accommodates  non-trivial initial conditions for the scalar field at the beginning of inflation as well as a finite duration of inflation.  The correspondence between the two 
  is encoded in the average field size  approaching the equilibrium value,
 \begin{equation}    \langle \overline {\Phi}^2 \rangle \rightarrow      {3\over 8\pi^2} \,{H^4 \over m^2}   \label {limit}  \end{equation}
 in the limit of infinitely long inflation.  The resulting particle abundances agree in this case.}
    \item{pre-inflationary initial conditions and finite duration of inflation make a crucial impact on the eventual particle abundance. The $\langle \overline {\Phi}^2 \rangle $ equilibrium value is approached very slowly: it takes about $H^2/m^2$ Hubble times for a free scalar 
    and $1/\sqrt{\lambda}$ Hubble times for a feebly interacting scalar to reach it. This corresponds to an ``exponentially'' long inflation in the sense that the required number of $e$-folds is exponentially large. Therefore, on a shorter time scale, the average field value at the end of inflation is often determined   by the pre-inflationary initial condition,
    $$  \langle \overline {\Phi}^2 \rangle  \sim  \langle \overline {\Phi}^2 \rangle_0 \;. $$
    This can be very large: after all, the inflaton field value is trans-Planckian at this stage, so it would be naive to expect  $ \langle \overline {\Phi}^2 \rangle_0$ to be negligible fortuitously.  
    The unknown $ \langle \overline {\Phi}^2 \rangle_0$ as well as the total duration of inflation result in non-removable  uncertainty in the eventual relic abundance $Y$.
     The average long-wavelength field at the end of inflation $\overline {\Phi} = \sqrt{\langle \overline {\Phi}^2 \rangle}$  can vary between $H$ and  $M_{\rm Pl}$, which results in the $Y$-uncertainty of many orders of magnitude, e.g. at least 10 orders of magnitude for a free scalar. } 
        \item{the relic abundance of  particles produced  via inflation exhibits a universal scaling
       \begin{equation}
       Y \propto {\overline {\Phi}^2 \over m^{1/2 }_{\rm eff}} \times  \left(   H_R \over  m_{\rm eff}  \right)^\gamma \;,
       \label{scaling}
       \end{equation} 
       where $\gamma=0$ and $ 1/2$ 
   for the radiation and matter dominated epochs following inflation, respectively.  $\overline {\Phi}$ represents the scalar condensate at the end of inflation and $m_{\rm eff}$ is the $postinflationary$ effective mass, i.e. the bare mass $m$ for a free scalar and 
   $\sqrt{3\lambda}\, \overline {\Phi}$ for a feebly interacting scalar.  $H_R$ is the Hubble rate at reheating and 
    the factor $H_R /  m_{\rm eff} $ can be very small, representing   dilution of the produced particles in the matter-dominated epoch.
   
   The above result also applies to  a scalar field with 
   a small non-minimal coupling to gravity, $|\xi |  \ll 1$.  It does not affect $m_{\rm eff}$ and may only affect the equilibrium value of $\overline {\Phi}$ via an inflation-induced mass $\sqrt{12\xi} H$  in Eq.\,\ref{limit}  (see Sec.\,\ref{sub-xi}).
   However, as explained above, we treat $\overline {\Phi}$ as a free parameter.
            }
            \item{one can set a lower bound  on the abundance of particles produced via inflation. The evolution equation requires  $\overline {\Phi}$ to be at least of order $H$,
            $$    \overline {\Phi} \gtrsim {\cal O}(H) \;.   $$
             In the context of high scale inflation, this results  in a large amount of dark relics.
            Their abundance is consistent with observations only if the relics are {\it very light} and/or the reheating temperature is {\it very low}. The  corresponding constraints are given by Eqs.\,\ref{Phi-free-rad},\ref{matter-Phi},\ref{SY-mat},\ref{SY-rad}.
            For example, the existence of a  free stable scalar is allowed only    if its mass is far below an eV or  the reheating temperature is in the GeV range or below.            
            A small non-minimal coupling to gravity does not affect these results. }
   \item{inflationary particle production is suppressed  if the scalar attains a large effective mass during inflation. 
   This can be achieved via a direct scalar coupling to the inflaton or  significant non-minimal coupling to gravity.  
   However, such interactions lead  to efficient particle production during preheating, which reintroduces the problem.  }
    \end{itemize}

\section{Conclusion}

We have studied inflationary  particle production of free and feebly interacting scalars.  It can be analyzed using the Bogolyubov coefficient method or with the help of the Starobinsky stochastic approach.  
The standard Bogolyubov approach assumes the Bunch-Davies boundary conditions for the scalar field at the beginning of inflation, corresponding to the infinite past, and represents an idealized situation.
The Starobinsky formalism, on the other hand,  readily accommodates 
 non-trivial initial conditions as well as a finite duration of inflation. 
We find that the two approaches agree in the limit of infinitely long inflation, while the Starobinsky method is more appropriate for studying   realistic situations.

A  spectator scalar  field is expected to have  a non-zero   value at the start of inflation, in analogy with the inflaton field itself. If the spectator is light, its average field size approaches  the equilibrium value very slowly.
Therefore, for a finite duration of inflation,  the average field value at the end of inflation (the ``condensate'')    is often determined by the initial conditions rather than the asymptotic equilibrium value.
As a result, the eventual particle abundance is sensitive to unknown {\it pre-inflationary} initial conditions as well as to the duration of inflation. The consequent uncertainty in the relic abundance spans many orders of magnitude making predictions 
 all but impossible. 
 
 Nevertheless, it is possible to obtain a lower bound on the produced particle abundance. In the framework of high scale inflation, the amount of produced light  scalars, with masses below the inflationary Hubble rate,  
 is very large. If such scalars are stable, their abundance is bounded by the abundance of dark matter. We find that this constraint is satisfied only if the particles are extremely light    and/or the reheating temperature is very low. For example, a  free stable scalar 
 must have a sub-eV mass or the reheating temperature has to be in the GeV range or below.
   
Our results also apply if the scalar has a small non-minimal coupling to gravity, $|\xi| \ll 1$, which, for instance,  can be generated via radiative corrections.  It creates an effective mass term during inflation that    affects the asymptotic value of 
the field condensate. However,  since we treat the condensate as a free variable, the effect  of a small $\xi$ is insignificant.
We find a universal scaling behaviour of the particle abundance produced via inflation (\ref{scaling}), which applies to free and feebly interacting scalars with zero or small non-minimal coupling to gravity.
 
 Inflationary particle production can be suppressed if the spectator attains a large inflation-induced mass, for instance, via a coupling to the inflaton or scalar curvature $R$. However, this leads to efficient postinflationary
 particle production during the inflaton oscillation epoch. As a result, the problem of dark relic ``overproduction'' is reintroduced under a different guise. 
  
 Our findings have important implications for non-thermal dark matter model building. Indeed, in the case of very weakly interacting dark matter, its  abundance is additive and thus determined by all of the production mechanisms combined. Since 
 gravitational particle production is always present and  particularly efficient during inflation, it generates a ubiquitous  background and 
 must be accounted for. This problem is exacerbated  by the quantum gravity effects, which generate higher-dimensional  operators responsible for particle production during preheating.


\begin{thebibliography}{99}
  
  
\bibitem{Mukhanov:2005sc}
V.~Mukhanov,
``Physical Foundations of Cosmology,'' Cambridge University Press, 2005;
doi:10.1017/CBO9780511790553.
  
  
   
\bibitem{Parker:1969au}
L.~Parker,
Phys. Rev. \textbf{183}, 1057-1068 (1969);
A.~A.~Grib and S.~G.~Mamaev,
Yad. Fiz. \textbf{10}, 1276-1281 (1969);
Y.~B.~Zeldovich and A.~A.~Starobinsky,
Zh. Eksp. Teor. Fiz. \textbf{61}, 2161-2175 (1971)

\bibitem{Parker:1971pt}
L.~Parker,
Phys. Rev. D \textbf{3}, 346-356 (1971)
[erratum: Phys. Rev. D \textbf{3}, 2546-2546 (1971)].

\bibitem{Grib:1976pw}
S.~G.~Mamaev, V.~M.~Mostepanenko and A.~A.~Starobinsky,
Zh. Eksp. Teor. Fiz. \textbf{70}, 1577-1591 (1976);
A.~A.~Grib, S.~G.~Mamaev and V.~M.~Mostepanenko,
Gen. Rel. Grav. \textbf{7}, 535-547 (1976).

      
\bibitem{Ford:1986sy}
L.~H.~Ford,
Phys. Rev. D \textbf{35}, 2955 (1987).

\bibitem{Ford:2021syk}
L.~H.~Ford,
Rept. Prog. Phys. \textbf{84}, no.11, 116901 (2021).



\bibitem{Kolb:2023ydq}
E.~W.~Kolb and A.~J.~Long,
Rev. Mod. Phys. \textbf{96}, no.4, 045005 (2024).






\bibitem{Starobinsky:1980te}
A.~A.~Starobinsky,
Phys. Lett. B \textbf{91} (1980) 99-102.


\bibitem{Guth:1980zm}
A.~H.~Guth,
Phys. Rev. D \textbf{23} (1981) 347-356.

\bibitem{Linde:1981mu}
A.~D.~Linde,
Phys. Lett. B \textbf{108} (1982) 389-393; Phys. Lett. B \textbf{129} (1983), 177-181.




\bibitem{Ema:2015dka}
Y.~Ema, R.~Jinno, K.~Mukaida and K.~Nakayama,
JCAP \textbf{05}, 038 (2015);
Y.~Ema, R.~Jinno, K.~Mukaida and K.~Nakayama,
Phys. Rev. D \textbf{94}, no.6, 063517 (2016).



\bibitem{Garny:2015sjg}
M.~Garny, M.~Sandora and M.~S.~Sloth,
Phys. Rev. Lett. \textbf{116}, no.10, 101302 (2016).


\bibitem{Mambrini:2021zpp}
Y.~Mambrini and K.~A.~Olive,
Phys. Rev. D \textbf{103}, no.11, 115009 (2021).

\bibitem{Lebedev:2022ljz}
O.~Lebedev and J.~H.~Yoon,
JCAP \textbf{07}, no.07, 001 (2022).









  
\bibitem{Bogolyubov:1958se}
N.~N.~Bogolyubov,
Sov. Phys. JETP \textbf{7}, 41-46 (1958)
JINR-R-94.


   
\bibitem{Starobinsky:1986fx}
A.~A.~Starobinsky,
Lect. Notes Phys. \textbf{246}, 107-126 (1986).
   
   
   
   \bibitem{Lebedev:2022cic}
O.~Lebedev,
JCAP \textbf{02}, 032 (2023).
   
   
   
   
   
\bibitem{Chernikov:1968zm}
N.~A.~Chernikov and E.~A.~Tagirov,
Ann. Inst. H. Poincare A Phys. Theor. \textbf{9}, 109 (1968). 
   
   
\bibitem{Bunch:1978yq}
T.~S.~Bunch and P.~C.~W.~Davies,
Proc. Roy. Soc. Lond. A \textbf{360}, 117-134 (1978).
   
   
   
\bibitem{Chung:1998zb}
D.~J.~H.~Chung, E.~W.~Kolb and A.~Riotto,
Phys. Rev. D \textbf{59}, 023501 (1998).
   
   
\bibitem{Chung:1998ua}
D.~J.~H.~Chung, E.~W.~Kolb and A.~Riotto,
Phys. Rev. Lett. \textbf{81}, 4048-4051 (1998). 
   
    
\bibitem{Kuzmin:1998kk}
V.~Kuzmin and I.~Tkachev,
Phys. Rev. D \textbf{59}, 123006 (1999).
   
   
   
   
\bibitem{Jenks:2024fiu}
L.~Jenks, E.~W.~Kolb and K.~Thyme,
[arXiv:2410.03938 [hep-ph]].

   
   
\bibitem{Buchbinder:2017lnd}
I.~L.~Buchbinder, S.~D.~Odintsov and I.~L.~Shapiro,
``Effective Action in Quantum Gravity,''
Routledge, 2017,
ISBN 978-0-203-75892-2.
   
   
   
   
   
     \bibitem{diff-eq}
      Valentin  Zaitsev and  Andrei  Polyanin,
      ``Handbook of Exact Solutions for Ordinary Differential Equations'', 
      Chapman and Hall/CRC, 2002.
     
        
   
   
   
\bibitem{master-th}
Julia Rantamaki, ``Application of the stochastic formalism for spectator scalars during inflation'', Master's Thesis, University of Jyvaskyla,
https://www.finna.fi/Record/jyx.123456789\_92074?imgid=1   
   
   
\bibitem{Grain:2017dqa}
J.~Grain and V.~Vennin,
JCAP \textbf{05}, 045 (2017).
   
   
\bibitem{Cable:2020dke}
A.~Cable and A.~Rajantie,
Phys. Rev. D \textbf{104}, no.10, 103511 (2021).
   
   
\bibitem{Kaneta:2023kfv}
K.~Kaneta and K.~y.~Oda,
JCAP \textbf{10}, 048 (2023).
   
   
\bibitem{Hannestad:2004px}
S.~Hannestad,
Phys. Rev. D \textbf{70}, 043506 (2004).

   
   
   
\bibitem{Lebedev:2021xey}
O.~Lebedev,
Prog. Part. Nucl. Phys. \textbf{120}, 103881 (2021).
   
   
\bibitem{Cosme:2023xpa}
C.~Cosme, F.~Costa and O.~Lebedev,
Phys. Rev. D \textbf{109}, no.7, 075038 (2024).
   
   
\bibitem{Lebedev:2024vor}
O.~Lebedev, A.~P.~Morais, V.~Oliveira and R.~Pasechnik,
[arXiv:2410.21874 [hep-ph]].
   
   
   
\bibitem{Starobinsky:1994bd}
A.~A.~Starobinsky and J.~Yokoyama,
Phys. Rev. D \textbf{50}, 6357-6368 (1994).
   
  
  
\bibitem{Peebles:1999fz}
P.~J.~E.~Peebles and A.~Vilenkin,
Phys. Rev. D \textbf{60}, 103506 (1999).
  
  
  
    
\bibitem{Markkanen:2018gcw}
T.~Markkanen, A.~Rajantie and T.~Tenkanen,
Phys. Rev. D \textbf{98}, no.12, 123532 (2018).

  
  
  
\bibitem{Arcadi:2019oxh}
G.~Arcadi, O.~Lebedev, S.~Pokorski and T.~Toma,
JHEP \textbf{08}, 050 (2019).
  
  
  
\bibitem{Bassett:1997az}
B.~A.~Bassett and S.~Liberati,
Phys. Rev. D \textbf{58}, 021302 (1998)
[erratum: Phys. Rev. D \textbf{60}, 049902 (1999)].

  
\bibitem{Kofman:1997yn}
L.~Kofman, A.~D.~Linde and A.~A.~Starobinsky,
Phys. Rev. D \textbf{56}, 3258-3295 (1997).
  
\bibitem{Greene:1997fu}
P.~B.~Greene, L.~Kofman, A.~D.~Linde and A.~A.~Starobinsky,
Phys. Rev. D \textbf{56}, 6175-6192 (1997).
  
\bibitem{Lebedev:2021tas}
O.~Lebedev, F.~Smirnov, T.~Solomko and J.~H.~Yoon,
JCAP \textbf{10}, 032 (2021).

   
\bibitem{Dolgov:1989us}
A.~D.~Dolgov and D.~P.~Kirilova,
Sov. J. Nucl. Phys. \textbf{51}, 172-177 (1990).

\bibitem{Traschen:1990sw}
J.~H.~Traschen and R.~H.~Brandenberger,
Phys. Rev. D \textbf{42}, 2491-2504 (1990).

\bibitem{Ichikawa:2008ne}
K.~Ichikawa, T.~Suyama, T.~Takahashi and M.~Yamaguchi,
Phys. Rev. D \textbf{78}, 063545 (2008).


\bibitem{Peskin:1995ev}
M.~E.~Peskin and D.~V.~Schroeder,
``An Introduction to quantum field theory,''
Addison-Wesley, 1995,
ISBN 978-0-201-50397-5.

   
   
   
\bibitem{Khlebnikov:1996mc}
S.~Y.~Khlebnikov and I.~I.~Tkachev,
Phys. Rev. Lett. \textbf{77}, 219-222 (1996).

   
   
   
   
   


\bibitem{Ema:2017loe}
Y.~Ema, M.~Karciauskas, O.~Lebedev and M.~Zatta,
JCAP \textbf{06}, 054 (2017).



\bibitem{Lebedev:2022vwf}
O.~Lebedev, T.~Solomko and J.~H.~Yoon,
JCAP \textbf{02}, 035 (2023).
   
   
   
   
   
   
\bibitem{Kaneta:2022gug}
K.~Kaneta, S.~M.~Lee and K.~y.~Oda,
JCAP \textbf{09}, 018 (2022).
   
   
\bibitem{Garcia:2022vwm}
M.~A.~G.~Garcia, M.~Pierre and S.~Verner,
Phys. Rev. D \textbf{107}, no.4, 043530 (2023).
   
   
\bibitem{Dvali:2022vzz}
G.~Dvali and L.~Eisemann,
Phys. Rev. D \textbf{106}, no.12, 125019 (2022).
   
   
\bibitem{Koutroulis:2023fgp}
F.~Koutroulis, O.~Lebedev and S.~Pokorski,
JHEP \textbf{04}, 027 (2024).
   
\bibitem{Chung:2011ck}
D.~J.~H.~Chung, L.~L.~Everett, H.~Yoo and P.~Zhou,
Phys. Lett. B \textbf{712}, 147-154 (2012).
   
\bibitem{Garcia:2025rut}
M.~A.~G.~Garcia, W.~Ke, Y.~Mambrini, K.~A.~Olive and S.~Verner,
[arXiv:2502.20471 [hep-ph]].
   
   
   
   
   
   \end{thebibliography}
\end{document}